\documentclass[a4paper,11pt]{article}
\usepackage{jheppub}\usepackage{lineno}
\usepackage{amsthm,amsmath,amssymb,amsfonts}
\usepackage{mathrsfs,bm}
\usepackage{slashed,braket,tensor,cancel}
\usepackage[dvipsnames]{xcolor}
\usepackage{tikz,tikz-cd,tkz-euclide}
\usepackage[compat=1.1.0]{tikz-feynman}
\usetikzlibrary{cd}\usetikzlibrary{calc}\usetikzlibrary{patterns}
\usetikzlibrary{feynman}\tikzfeynmanset{warn luatex=false}\usetikzlibrary{arrows.meta}
\allowdisplaybreaks[4]
\def\eq{\eqref}\def\nn{\nonumber}

\def\c.c.{\mathrm{c.c.}}

\def\d{\mathrm{d}}\def\i{\mathrm{i}}
\newcommand{\e}{\mathrm{e}}\newcommand{\p}{\partial} 
\def\0{{(0)}}\def\1{{(1)}}\def\2{{(2)}}\def\3{{(3)}}\def\4{{(4)}}

\def\qrq{\quad\Rightarrow\quad}\def\qlq{\quad\Leftrightarrow\quad}
\def\qaq{\quad\text{and}\quad}
\def\qwq{\quad\text{with}\quad}
\def\pl{\text{\pounds}}
\def\wt{\widetilde}\def\wh{\widehat}

\newcommand{\SO}{\mathrm{SO}}

\DeclareMathOperator{\tr}{tr}

\def\mn{{\mu\nu}}\def\rs{{\rho\sigma}}
\def\a{\alpha}\def\b{\beta}\def\g{\gamma}
\def\de{\delta}\def\ep{\epsilon}\def\th{\theta}
\def\k{\kappa}\def\l{\lambda}
\def\m{\mu}\def\n{\nu}
\def\r{\rho}\def\s{\sigma}
\def\vp{\varphi}
\def\De{\Delta}\def\Th{\Theta}\def\L{\Lambda}\def\Si{\Sigma}

\def\ca{\mathcal{A}}\def\cb{\mathcal{B}}

\def\cf{\mathcal{F}}
\def\ch{\mathcal{H}}
\def\ci{\mathcal{I}}
\def\ck{\mathcal{K}}\def\cl{\mathcal{L}}

\def\co{\mathcal{O}}
\def\ct{\mathcal{T}}

\def\cy{\mathcal{Y}}\def\cz{\mathcal{Z}}

\def\cA{\mathcal{A}}\def\cB{\mathcal{B}}

\def\cF{\mathcal{F}}

\def\cL{\mathcal{L}}

\def\cR{\mathcal{R}}
\def\cT{\mathcal{T}}

\def\cX{\mathcal{X}}
\def\cY{\mathcal{Y}}\def\cZ{\mathcal{Z}}

\def\R{\mathbb{R}}

\title{\boldmath Covariant variation and its applications}
\author{Wen-Bin Liu}\author{and Jiang Long}
\affiliation{School of Physics, Huazhong University of Science and Technology,\\
Luoyu Road 1037, Wuhan, Hubei, 430074, China}
\emailAdd{liuwenbin0036@hust.edu.cn, longjiang@hust.edu.cn}
\abstract{We define a covariant variation of tensor fields by combining its Lie derivative with the metric variation. This operator preserves the metric, contractions, and Hodge duality, but its commutator is not closed due to an anomaly. We derive its algebraic and geometric properties, and compare it with the Kosmann derivative. Combining the covariant variation with Kosmann derivative gives total covariant variation for the fields with both spacetime and Lorentz structure, all of which belong to the metric Lie derivative. Moreover, we introduce families of extended operators which contain the affine connection, Lie derivative, and covariant variation. From the anomaly of the covariant variation along the superrotation, an electromagnetic helicity flux appears at null hypersurfaces in four dimensions. We also apply the covariant variation and its anomaly to tensor fields in arbitrary spacetime dimensions, and especially focus on the $p$-forms in $d=2p+2$ dimensions.}

\begin{document}
\maketitle
\flushbottom

\section{Introduction}
The Lie derivative is the canonical infinitesimal action of a vector field on ordinary spacetime tensors. It preserves tensor products and contractions, but it does not preserve a chosen metric unless the generating vector field is Killing. Consequently, for a generic vector field, Lie transport does not commute with raising and lowering indices or with the Hodge star. If we want to consider a quantum field theory under some fixed background, only isometries act as symmetries, and generic diffeomorphisms instead relate different background configurations. 

The issue appears concretely in radiative Maxwell theory at future null infinity. The flux associated with a general superrotation does not act on the angular gauge potential by the usual Lie derivative. Instead, its action contains a compensating term determined by the variation of the metric on the celestial sphere. Moreover, the commutator of two such transformations contains a local rotation of the radiative polarization, which is generated by the electromagnetic helicity flux, which concerns the electromagnetic duality \cite{1965AmJPh..33..958C,Deser:1976iy,Afanasev:1995vh,Fernandez-Corbaton:2012ikh,Bliokh:2012zr,Elbistan:2016khp,Agullo:2018nfv,Liu:2023qtr,Liu:2024rvz}. Analogous helicity operators have appeared for the gravity \cite{Henneaux:2004jw,Liu:2023gwa,Liu:2024nkc}, higher-spin fields \cite{Liu:2023jnc}, and fermions \cite{Guo:2024qzv}. Both electromagnetic and gravitational helicity fluxes are interesting potential observables \cite{Long:2024yvj,Heng:2025kmr}, which generate nonlinear spin memory effects \cite{Seraj:2022qyt,Oblak:2023axy,Maleknejad:2023nyh}. Moreover, these fluxes can be derived from some specific topological terms \cite{Nieh:1981ww,Nieh:1981xk,Nieh:2007zz,Liu:2024rvz,Long:2025fbb}.

The above observations motivate a systematic study of the metric-compensated transformation itself. Let $X$ be a vector field on a pseudo-Riemannian manifold $(M,g)$, and let $\rho$ denote the natural slotwise action of $\operatorname{End}(TM)$ on an ordinary tensor bundle. We define
\begin{align}
\Delta_X=\mathcal L_X+\rho(S_X)=\nabla_X+\rho(A_X),
\qquad
S_X=\frac12 g^{-1}\mathcal L_Xg,
\qquad
(A_X)_{\mu\nu}=\nabla_{[\mu}X_{\nu]}.
\end{align}
The two expressions in the first equation agree for the Levi-Civita connection. By construction, $\Delta_X$ is a derivation of the tensor algebra, commutes with contractions, annihilates the metric and volume form, and commutes with the Hodge star. We call it the covariant variation. Here covariant means metric-compatible on ordinary spacetime tensors, but it does not mean covariant with respect to an independent internal gauge bundle.

The price of metric compatibility is non-closure for generic vector fields. We find
\begin{align}
[\Delta_X,\Delta_Y]-\Delta_{[X,Y]}
=\rho\bigl(-[S_X,S_Y]\bigr).
\end{align}
Thus, the non-closure acts through the tensor representation as an infinitesimal orthogonal transformation. It vanishes on the isometry algebra and, more generally, whenever either vector field is conformal Killing. The operator Jacobi identity gives the corresponding Bianchi-type identity. We use the word anomaly as shorthand for this classical non-closure endomorphism, since it is closely related to the chiral anomaly \cite{Adler:1969gk,Bell:1969ts}.

Geometrically, $\Delta_X$ is the metric Lie derivative \cite{Bourguignon:1992sp}, or more generally reductive Lie derivative \cite{Godina:2003tc} induced by a metric-dependent lift of $X$ to the orthonormal frame bundle \cite{Giotopoulos:2025weu,Dalak:2026zbz}. The same lift acts as the Kosmann derivative on Lorentz tensors and spinors \cite{Lichnerowicz1963,Kosmann:1971ugf,Jackiw:1979ub,Ortin:2002qb,Obukhov:2006ge,Fatibene:2009,Jacobson:2015uqa,Prabhu:2015vua,Aneesh:2020fcr,Elgood:2020svt}. An internal gauge bundle is not canonically identified with $TM$, and we need vielbein to relate internal Lorentz components with spacetime tensor components. It is known that these operators suffer an anomaly under the commutator, but there is no exploration yet about the physical meaning of this anomaly in the literature. We found, in certain cases, it gives helicity flux or spin charge of the massless or massive spinning field \cite{Liu:2025oom}. 

We apply the covariant variation to null hypersurfaces. In two transverse dimensions, every antisymmetric endomorphism is proportional to the area form, so the non-closure acts as a helicity rotation. In higher dimensions, it becomes an $\mathfrak{so}(d-2)$-valued fiber rotation on the radiative data. This statement does not by itself establish a flux algebra. The fluxes exist only after specifying an action or at least a symplectic structure.

The paper is organized as follows. Section~\ref{sec:motivation} derives the covariant variation from the Maxwell superrotation flux and distinguishes its bulk, intrinsic-boundary, and half-density realizations. In section~\ref{sec:general-properties}, we establish its algebraic properties, commutators, geometric interpretation, and action on differential forms. More extended differential operators are discussed in the following section. In section~\ref{sec:null-boundary-applications}, we consider some applications, including the electromagnetic field near horizon and arbitrary tensor fields in general dimensions. Technical details concerning higher rank differential forms, affine connections, gauge-covariant lifts, and curvature, as well as the Jacobi identities, are collected in the appendices.

\section{Motivation of covariant variation}\label{sec:motivation}
\subsection{Superrotation and covariant variation}\label{sec SR CV}
We now introduce some conventions. For flat spacetime, it is convenient to describe the future null infinity $\ci^+$ in retarded coordinates $(u,r,x^A)$ with $x^A=(\th,\phi)$ for the sphere, and the metric reads 
\begin{align}
\d s^2=-\d u^2-2\d u\d r+r^2(\d\th^2+\sin^2\th \d\phi^2).\label{metric}
\end{align}
The retarded coordinates are related to the Cartesian coordinates through
\begin{align}
x^\mu=u\bar m^\m+r n^\mu \qwq \bar m^\m=(1,0)=\frac{1}{2}(n^\mu-\bar n^\mu)
\end{align} 
where we have already defined two null vectors $n^\mu$ and $\bar n^\mu$ 
\begin{align}
n^\mu=(1,n^i),\qquad \bar n^\mu=(-1,n^i) \label{null}
\end{align}
with $n^i$ the normal vector of the unit sphere. We further introduce 
\begin{align}
m^\m=\frac{1}{2}(n^\mu+\bar n^\mu)=(0,n^i) \qaq Y^A_\m=-\nabla^An_\mu,
\end{align}
where the covariant derivative $\nabla_A$ is adapted to the standard metric on the sphere
\begin{align}
\d s^2_{S^2}=\d\th^2+\sin^2\th \d\phi^2\equiv \g_{AB}\d x^A \d x^B.
\end{align}
The six conformal Killing vector fields on $S^2$ are encoded by $Y^A_\mn=Y^A_\m n_\n-Y^A_\n n_\m$ whose properties are summarized in the appendix of \cite{Liu:2023gwa}. The integral measure on $\ci^+$ is denoted by
\begin{align}
\int \d u\d\Omega\equiv\int_{-\infty}^{\infty}\d u\int_0^{2\pi}\d\phi\int_0^\pi\d\th\,\sin\th.
\end{align}

In \cite{Liu:2023qtr}, we considered the action of the superrotation \cite{Campiglia:2014yka,Campiglia:2015yka,Compere:2018ylh} on a Maxwell field in the flat spacetime. The superrotation is generated by
\begin{align}
\xi_{\cal Y}=\frac{1}{2}u\nabla\cdot {\cal Y}\partial_u-\frac{1}{2}r\nabla\cdot {\cal Y}\partial_r+\frac{u }{4}\nabla^2\nabla\cdot {\cal Y}\p_r+({\cal Y}^A-\frac{u}{2r}\nabla^A\nabla\cdot {\cal Y})\partial_A+\cdots,\label{xiy}
\end{align}
where $\cy^A$ is a smooth vector field on the sphere. For our purpose, we only need the field-independent part. From the stress tensor, we can construct the superrotation flux
\begin{align}
\cf_\cy&=\int_{\ci^+}(\d^3x)_\m {T^\m}_{\n} \xi_{\cal Y}^\n\nn\\
&=\int \d u\d\Omega\,\Big[\frac{1}{2}u\nabla\cdot \cY\dot \cA_A\dot \cA^A+ \cY^A \dot{\cA}^B\nabla^C\cA^DP_{ABCD}\Big],\label{def cf cy}
\end{align}
where $\ca_A$ is the leading-order Maxwell field of large-$r$ expansion, i.e., $A_A=\ca_A+O(1/r)$, and we have introduced
\begin{align}
P_{ABCD}=\gamma_{AB}\gamma_{CD}+\gamma_{AC}\gamma_{BD}-\gamma_{AD}\gamma_{BC}.\label{Pabcd}
\end{align}

With the help of the fundamental commutator between the boundary fields
\begin{align}
[\cA_{A}(u,\Omega),\dot{\cA}_{B}(u',\Omega')]&=\frac{\i}{2}\gamma_{AB}\delta(u-u')\delta(\Omega-\Omega'),\label{fund comm}
\end{align}
we can compute the commutator between the superrotation flux and the boundary field
\begin{align}
[\i\cf_\cy, \cA_{A}]=\frac{1}{2}u\nabla\cdot \cY\dot\ca_A+{\cal Y}^B\nabla_B\cA_A+\cA^B\nabla_{[A}{\cal Y}_{B]}+\frac{1}{2}\ca_A\nabla\cdot {\cal Y}.
\end{align}
This should be compared with the classical boundary variation
\begin{align}
\delta_{\cal Y} \cA_A=\frac{1}{2}u\nabla\cdot \cY\dot\ca_A+{\cal Y}^B\nabla_B\cA_A+\cA_B\nabla_{A}{\cal Y}^{B},
\end{align}
which comes from the bulk Lie derivative
\begin{align}
\cl_{\xi_{\cal Y}}A_A=\delta_{\cal Y} \cA_A+O(1/r).\label{clxiyAA}
\end{align}
However, these two do not agree, and the difference is exactly the boundary metric variation
\begin{align}
\delta_{\cal Y}\g_{AB}=\lim_{r\to\infty} r^{-2}\cl_{\xi_{\cal Y}}\eta_{AB}=\nabla_A {\cal Y}_B+\nabla_B{\cal Y}_A-\g_{AB}\nabla\cdot {\cal Y}\equiv\Th_{AB}({\cal Y}).\label{deygab}
\end{align}
This inspires us to propose a covariant variation
\begin{align}
\Delta_{\cal Y} \cA_A&=\delta_{\cal Y} \cA_A-\frac{1}{2}\delta_{\cal Y} \gamma_{AB} \cA^B\\
&=\frac{1}{2}u\nabla\cdot {\cal Y}\dot\cA_A+{\cal Y}^B\nabla_B\cA_A+\cA^B\nabla_{[A}{\cal Y}_{B]}+\frac{1}{2}\ca_A\nabla\cdot {\cal Y}
\end{align}
which matches the action of superrotation flux on $\ca_A$
\begin{align}
[\i\cf_{\cal Y},\ca_A]=\Delta_{\cal Y} \cA_A.
\end{align}

The boundary metric variation \eq{deygab} vanishes only for conformal Killing vectors on the sphere, which corresponds to the Lorentz transformations in the bulk, while the corresponding covariant variation vanishes identically
\begin{align}
\Delta_{\cal Y}\g_{AB}=\delta_{\cal Y}\g_{AB}-\frac{1}{2}\Th_A^C({\cal Y})\g_{CB}-\frac{1}{2}\Th_B^C({\cal Y})\g_{CA}=0
\end{align}
for any vector field ${\cal Y}^A$. That is, we introduced a kind of variation of matter field which contains the change of background metric, and thus the same rule applies to the metric tensor yielding zero. The underlying logic is as follows: computing the commutator from \eq{fund comm} usually assumes a fixed boundary metric $\g_{AB}$, but the superrotation changes it, and therefore, we need to subtract the boundary metric variation from the classical variation $\delta_\cy\ca_A$ so that the resulting $\Delta_\cy\ca_A$ agrees with $[\i\cf_{\cal Y},\ca_A]$.

Moreover, a surprising helicity flux emerges from the quantum commutator of two superrotation fluxes \cite{Liu:2023qtr}
\begin{align}
[\cf_{\cal Y},\cf_{\cal Z}]=\i\cf_{[{\cal Y},{\cal Z}]}+\i\co_{o({\cal Y},{\cal Z})},\label{cmycmz}
\end{align}
where $o({\cal Y},{\cal Z})$ is defined as
\begin{align}
o({\cal Y},{\cal Z})=\frac{1}{4}\ep^{AB}\Th_{AC}({\cal Y})\Th^C_B({\cal Z}).
\end{align}
This helicity flux
\begin{align}
\co_h=\int\d u\d\Omega\,h(\Omega)\dot\ca_A\ca_B\ep^{BA}\label{ohscri}
\end{align}
concerns the bulk electromagnetic duality as well as the second Chern character \cite{Liu:2024rvz}. For constant $h$, it measures the number difference between particles with opposite helicities, which can be seen from its expression in the momentum space\footnote{For the bulk Maxwell field, we assume the following mode expansion
\begin{align}
  A_\mu(x)=\sum_{\a=\pm}\int \widetilde{\d p}[\epsilon^{*\alpha}_\mu(\bm p)a_{\alpha}(\bm p)\e^{\i p\cdot x}+\epsilon^{\alpha}_\mu(\bm p)a^\dagger_{\alpha}(\bm p)\e^{-\i p\cdot x}] \qwq \widetilde{\d p}=\frac{\d^3p}{(2\pi)^32\omega_{\bm p}}.\nn
\end{align}
}
\begin{align}
\co_{h=1}=\int \widetilde{\d p}\,[a_+^{\dagger}(\bm p)a_+(\bm p)-a_-^{\dagger}(\bm p)a_-(\bm p)].
\end{align}
Moreover, we obtained analogous results in various theories, including the gravitational theory \cite{Liu:2023gwa}, the higher spin theory \cite{Liu:2023jnc,Liu:2024nkc}, and even in the fermionic theory \cite{Guo:2024qzv}.

The appearance of helicity flux in \eq{cmycmz} is actually the consequence of non-closure of the covariant variation
\begin{align}
[\Delta_{\cal Y},\Delta_{\cal Z}]\cA_A&=\Delta_{[{\cal Y},{\cal Z}]}\cA_A-\frac{1}{4}(\delta_{\cal Y}\g_{AB}\delta_{\cal Z}\g_{CD}-\delta_{\cal Z}\g_{AB}\delta_{\cal Y}\g_{CD})\g^{BC}\cA^D.
\end{align}
The last term measures the non-closure of classical commutator and results in
\begin{align}
-\frac{1}{4}(\delta_{\cal Y}\g_{AB}\delta_{\cal Z}\g_{CD}-\delta_{\cal Z}\g_{AB}\delta_{\cal Y}\g_{CD})\g^{BC}\cA^D=-o({\cal Y},{\cal Z})\epsilon_{AB}\cA^B,
\end{align}
which exactly agrees with the action of the emerging helicity flux on the Maxwell field
\begin{align}
-o({\cal Y},{\cal Z})\epsilon_{AB}\cA^B=[\i\co_{o({\cal Y},{\cal Z})},\ca_A].\label{DYDZ}
\end{align}

If we define
\begin{align}
\ca_{\cY,\cZ}\ca_A=[\Delta_{\cal Y},\Delta_{\cal Z}]\cA_A-\Delta_{[{\cal Y},{\cal Z}]}\cA_A\equiv (\ca_{\cY,\cZ})_{A}{}^B\ca_B,
\end{align}
then the non-closure endomorphism satisfies the following Bianchi-type identity
\begin{align}
\sum_{\rm cyclic}\big[\Delta_\cX(\ca_{\cY,\cZ})_{A}{}^B+(\ca_{\cX, [{\cal Y}, {\cal Z}]})_{A}{}^B\big]=0.
\end{align} 
Here ``cyclic'' denotes cyclic permutations of $\cX,{\cal Y},{\cal Z}$. Moreover, since
\begin{align}
(\ca_{\cX,{\cal Y}})_{AB}=-o(\cX,{\cal Y})\ep_{AB},
\qquad \Delta_\cX\ep_{AB}=0,
\end{align}
the preceding identity is therefore equivalent to
\begin{align}
\sum_{\rm cyclic}\big[\cX^A\nabla_Ao({\cal Y},{\cal Z})+o(\cX, [{\cal Y}, {\cal Z}])\big]=0.
\end{align}

\subsection{Bulk covariant variation}
Both $\delta_\cY\ca_A$ and $\delta_\cY\g_{AB}$ are derived from the bulk Lie derivative, and the quantum flux $\cf_\cY$ also comes from the bulk stress tensor. Therefore, it is necessary to investigate the bulk covariant variation.

Following the same spirit, we can define a bulk version of covariant variation
\begin{subequations}\label{1.7a}
\begin{align}
\Delta_X V^\m &=\cl_X V^\m-\frac{1}{2}\cl_X g^{\m\n}V_\n,\\
\Delta_X A_\m &=\cl_X A_\m-\frac{1}{2}\cl_X g_{\m\n}A^\n.\label{DelxiA_m}
\end{align}
\end{subequations}
Introducing 
\begin{align}
(S_X)^\m_\n \equiv \frac{1}{2}  g^{\m\rho} \mathcal{L}_Xg_{\r\n}=-\frac{1}{2} \mathcal{L}_X g^{\m\rho} g_{\r\n},
\end{align}
we can rewrite the covariant variation more transparently as
\begin{subequations}
\begin{align}
&\Delta_X V^\m=\cl_XV^\m+(S_X)^\m_\n V^\n,\\ 
&\Delta_X A_\m=\cl_XA_\m-(S_X)_\m^\n A_\n.
\end{align}
\end{subequations}
If the indices are symmetric, their relative order does not matter. If they are not symmetric, especially if they are antisymmetric, we need to be cautious. Since $S_X$ acts differently on vectors and covectors, with the action extended slotwise to arbitrary tensors, we obtain the concise expression
\begin{align}
\De_X=\cl_X+\r(S_X) \qwq S_X=\frac{1}{2}g^{-1}\cl_Xg.
\end{align}
One can explicitly write how the $\r(S_X)$ acts on a general rank-$(k,l)$ tensor $T$
\begin{align}
[\r(S_X)T]^{\mu_{1}\cdots\mu_{k}}_{\nu_{1}\cdots\nu_{l}}=\sum_{i=1}^{k}(S_X)^{\m_i}_\r T^{\mu_{1}\cdots\r\cdots\mu_{k}}_{\nu_{1}\cdots\nu_{l}}-\sum_{i=1}^{l}(S_X)_{\n_i}^\r T^{\mu_{1}\cdots\mu_{k}}_{\nu_{1}\cdots\r\cdots\nu_{l}},
\end{align}
and thus we have
\begin{align}
\Delta_XT^{\mu_{1}\cdots\mu_{k}}_{\nu_{1}\cdots\nu_{l}}=\cl_XT^{\mu_{1}\cdots\mu_{k}}_{\nu_{1}\cdots\nu_{l}}+\sum_{i=1}^{k}(S_X)^{\m_i}_\r T^{\mu_{1}\cdots\r\cdots\mu_{k}}_{\nu_{1}\cdots\nu_{l}}-\sum_{i=1}^{l}(S_X)_{\n_i}^\r T^{\mu_{1}\cdots\mu_{k}}_{\nu_{1}\cdots\r\cdots\nu_{l}}.\label{DeX T}
\end{align}
One can easily verify that the covariant variation annihilates the metric identically
\begin{align}
\Delta_X g_\mn=\cl_X g_\mn-\cl_X g_\mn=0.
\end{align}

For a Levi-Civita connection, we know
\begin{align}
2(S_X)_\mn=\cl_X g_\mn=\nabla_\m X_\n+\nabla_\n X_\m.
\end{align}
$S_X$ is the symmetric part of the covariant gradient $\nabla_\m X_\n$. After subtracting it, we can express the covariant variation in terms of covariant derivative as
\begin{align}
\Delta_X=\nabla_X+\r(A_X),
\end{align}
where $A_X$ is the antisymmetric part of the covariant gradient
\begin{align}
2(A_X)_\mn=(\d X^\flat)_\mn=\nabla_\m X_\n-\nabla_\n X_\m.
\end{align}
Here the musical maps $\flat$ and $\sharp$ lower and raise indices with the metric. We can explicitly write out
\begin{subequations}
\begin{align}
[\r(A_X)T]^{\mu_{1}\cdots\mu_{k}}_{\nu_{1}\cdots\nu_{l}}&=\sum_{i=1}^{k}(A_X)^{\m_i}{}_\r T^{\mu_{1}\cdots\r\cdots\mu_{k}}_{\nu_{1}\cdots\nu_{l}}-\sum_{i=1}^{l}(A_X)^\r{}_{\n_i} T^{\mu_{1}\cdots\mu_{k}}_{\nu_{1}\cdots\r\cdots\nu_{l}}\\
&=\sum_{i=1}^{k}(A_X)^{\m_i}{}_\r T^{\mu_{1}\cdots\r\cdots\mu_{k}}_{\nu_{1}\cdots\nu_{l}}+\sum_{i=1}^{l}(A_X)_{\n_i}{}^\r T^{\mu_{1}\cdots\mu_{k}}_{\nu_{1}\cdots\r\cdots\nu_{l}}.
\end{align}
\end{subequations}
Now \eqref{DeX T} may be rewritten as
\begin{align}
\Delta_XT^{\mu_{1}\cdots\mu_{k}}_{\nu_{1}\cdots\nu_{l}}=\nabla_XT^{\mu_{1}\cdots\mu_{k}}_{\nu_{1}\cdots\nu_{l}}+\sum_{i=1}^{k}(A_X)^{\m_i}{}_\r T^{\mu_{1}\cdots\r\cdots\mu_{k}}_{\nu_{1}\cdots\nu_{l}}+\sum_{i=1}^{l}(A_X)_{\n_i}{}^\r T^{\mu_{1}\cdots\mu_{k}}_{\nu_{1}\cdots\r\cdots\nu_{l}}.
\end{align}

The Lie derivative defines a representation of the Lie algebra of vector fields, i.e., it is closed under the commutator
\begin{align}  
[\cl_{X},\cl_{Y}]=\cl_{[X, Y]},\label{LXLY}
\end{align}
while the covariant variation is not closed in general. For any ordinary tensor $T$, define the classical non-closure endomorphism by
\begin{subequations}\label{general anomaly}
\begin{align}  
\ca_{X, Y}T&\equiv[\Delta_{X},\Delta_{Y}]T-\Delta_{[X, Y]}T,\\
(\ca_{X, Y})^\m{}_\n&=\frac{1}{4}\cl_X g^{\m\r}\cl_Yg_{\r\n}-(X\leftrightarrow Y)= -[S_X, S_Y]^\m{}_\n.
\end{align}
\end{subequations}
Because $S_X$ is symmetric, $\ca_{X,Y}$ is antisymmetric. It is a classical curvature-like failure of closure, and we call it anomaly as shorthand.

Applying this formula to the superrotation \eq{xiy} and taking the Minkowski metric as the background, one obtains
\begin{align}
\cl_{\xi_{\cal Y}}\eta_{\mu\nu}&=-(\nabla\cdot \cy+\frac{1}{2}\nabla^2\nabla\cdot \cy)n_\m n_\n-\gamma_{\mu\nu}\nabla\cdot \cy+\nabla_A\cy_B(Y_{\mu}^A Y_\nu^B+Y_\nu^AY_\mu^B)+\cdots,\label{dyemn}
\end{align}
where $\g_\mn=Y_\m^A Y_{\n A}$ and thus 
\begin{align}
\ca_{\xi_{\cal Y},\xi_\cz}A_\m&=r^{-1}o(\cY,\cZ)Y^A_\m\epsilon_{AB}\cA^B+O(r^{-2}).\label{re}
\end{align}
Converting to angular component, we reproduce \eq{DYDZ}
\begin{align}
\ca_{\xi_{\cal Y},\xi_\cz}A_A=-o(\cY,\cZ)\epsilon_{AB}\cA^B+O(1/r).
\end{align}
In fact, the expected bulk-to-boundary relation holds for covariant variation
\begin{align}
\Delta_{\xi_{\cal Y}}A_A=\Delta_\cY\ca_A+O(1/r),
\end{align}
analogous to the case of Lie derivative \eq{clxiyAA}.

\subsection{Intrinsic boundary covariant variation}\label{sec intrin}
The previous boundary covariant variation in section \ref{sec SR CV} is reduced from the bulk one, which can agree with the action of the flux, since the latter one is also constructed from the bulk stress tensor. On the other hand, we can define the intrinsic boundary covariant variation of the Maxwell field
\begin{subequations}
\begin{align}
\bar\Delta_{\cy}\ca_A&=\cl_{\cy}\ca_A-\frac{1}{2}\cl_\cy\g_{AB}\ca^B\\
&=\cy^B\nabla_B\cA_A+\cA^B\nabla_{[A}\cy_{B]}
\end{align}
\end{subequations}
with respect to the vector field $\cy=\cy^A\p_A$ on the sphere. This $\bar\Delta_{\cy}\ca_A$ does not agree with the action of $\cf_\cy$. Even though we can remove the term with similar form as the supertranslation flux\footnote{This is the convention we used before, since we aim to construct the flux algebra \cite{Liu:2022mne,Liu:2023jnc,Liu:2023gwa} including the supertranslation.}
\begin{align}
\cf_f=\int \d u\d\Omega\,f(u,\Omega)\dot \cA_A\dot \cA^A
\end{align}
i.e., the first term in \eqref{def cf cy}, the remaining term acts as\footnote{A similar weighted generalized Lie derivative occurs in the double field theory \cite{Hohm:2010pp,Aldazabal:2013sca}.}
\begin{align}
[\i\cf_\cy-\i\cf_{f=\frac{1}{2}u\nabla\cdot \cY}, \cA_{A}]={\cal Y}^B\nabla_B\cA_A+\cA^B\nabla_{[A}{\cal Y}_{B]}+\frac{1}{2}\ca_A\nabla\cdot {\cal Y}\equiv\wt\Delta_\cy\ca_A.\label{pure sph of bulk reduction}
\end{align}
Unlike the previous $\Delta_\cy$, this $\wt\Delta_\cy$ is purely on the sphere.
The extra divergence term is forced by the radiative symplectic form. To see
this directly, introduce
\begin{align}
(B,C)_\g\equiv\int\d\Omega\,B^AC_A.
\end{align}
Integration by parts on the closed sphere, together with the
antisymmetry of $\nabla_{[A}\cy_{B]}$, gives
\begin{align}
(B,\bar\Delta_{\cY}C)_\g=-(\bar\Delta_{\cY}B,C)_\g-(B,\psi_{\cY}C)_\g,
 \label{intrinsic-formal-adjoint}
\end{align}
where we introduced $\psi_\cy=\nabla_A\cy^A$. Equivalently, this is
\begin{align}
\bar\Delta_{\cY}^{\dagger} =-\bar\Delta_{\cY}-\psi_{\cY}.
 \label{skew-boundary-operator}
\end{align}
For $\wt\Delta_\cy$, we have
\begin{align}
(B,\wt\Delta_{\cY}C)_\g=-(\wt\Delta_{\cY}B,C)_\g \qlq \wt\Delta_{\cY}^{\dagger} =-\wt\Delta_{\cY}.
\end{align}
Recall the Maxwell radiative symplectic form 
\begin{align}
\Omega(\delta\cA,\delta\cA)=\int\d u\,
\delta\cA_A\wedge\partial_u\delta\cA^A. \label{radiative-symplectic-form}
\end{align}
Let $G$ be a real, $u$-independent angular operator and set $\delta_G\cA=G\cA$.  Assuming the same endpoint conditions in $u$ that
are used in defining the flux, a direct substitution into
\eqref{radiative-symplectic-form} yields
\begin{align}
\delta_G\Omega=0\quad\Longleftrightarrow\quad G^\dagger=-G.
\label{Hamiltonian-skew-condition}
\end{align}
Therefore, we need to demand either $\psi_\cy=0$ or correct $\bar\Delta_\cy$ with $\frac{1}{2}\psi_\cy$. The previous one excludes part of CKVs on the sphere, which has actually been proposed in \cite{Guo:2024qzv} for $\xi_\cy$, called magnetic superrotation.

There is a simple intrinsic interpretation of the corrected operator. Let $\widehat{\cA}_A$ be a one-form density of weight $1/2$
\begin{align}
\widehat{\cA}_A=(\sqrt{\g})^{1/2}\cA_A
\end{align}
rather than an
ordinary one-form. Since
\begin{align}
\mathcal{L}_{\cY}(\sqrt{\g})^{1/2}=\frac12\psi_{\cY}(\sqrt{\g})^{1/2}.
\end{align}
Applying the metric compensation only to the tensor index gives
\begin{align}
\widehat\Delta_{\cY}\widehat{\cA}_A&=\widehat\Delta_{\cY}[(\sqrt{\g})^{1/2}\cA_A]\equiv \cl_{\cY}[(\sqrt{\g})^{1/2}\cA_A]- (\sqrt{\g})^{1/2}(S_{\cY})_A^B\cA_B\\ 
&= (\sqrt{\g})^{1/2}({\cal Y}^B\nabla_B\cA_A+\cA^B\nabla_{[A}{\cal Y}_{B]}+\frac{1}{2}\ca_A\nabla\cdot {\cal Y})\nn\\
&\equiv(\sqrt{\g})^{1/2}\wt\Delta_{\cY}\cA_A. \label{half-density-interpretation}
\end{align}
Thus the superrotation flux realizes the intrinsic covariant variation on an independently weighted radiative quantity. If one insists on a weight-zero one-form and the operator $\bar\Delta_{\cY}$, one must either restrict to area-preserving vector fields or enlarge the phase space so that the boundary metric can vary. 

Finally, although $\bar\Delta_{\cy}\ca_A$ differs from $\Delta_{\cy}\ca_A$ or $\wt\Delta_{\cy}\ca_A$, the anomaly is the same, i.e.,
\begin{subequations}
\begin{align}
[\bar\Delta_{\cal Y},\bar\Delta_{\cal Z}]\cA_A-\bar\Delta_{[{\cal Y},{\cal Z}]}\cA_A&=-\frac{1}{4}(\cl_{\cal Y}\g_{AB}\cl_{\cal Z}\g_{CD}-\cl_{\cal Z}\g_{AB}\cl_{\cal Y}\g_{CD})\g^{BC}\cA^D\\
&=-o({\cal Y},{\cal Z})\epsilon_{AB}\cA^B,
\end{align}
\end{subequations}
since $\psi_{[\cY,\cZ]} =\cY[\psi_{\cZ}]-\cZ[\psi_{\cY}]$.

\section{Algebraic and geometric properties}\label{sec:general-properties}
This section establishes the derivation properties, commutators, geometric interpretation,
and action on differential forms of the covariant variation.

\subsection{Algebraic structure and geometric interpretation}
\label{sec:algebraic-geometric-meaning}
For fixed $X$, the operator $\Delta_X$ is an $\mathbb R$-linear derivation
of the tensor algebra.  In particular,
\begin{subequations}\label{Delta-algebraic-summary}
\begin{align}
 \Delta_Xf&=X[f],\label{Delta-on-function}\\
 \Delta_X(fT)&=f\Delta_XT+X[f]T,\label{Delta-field-Leibniz}\\
 \Delta_X(T\otimes U)
 &=\Delta_XT\otimes U+T\otimes\Delta_XU,\label{Delta-tensor-Leibniz}\\
 \Delta_X\operatorname{C}(T)
 &=\operatorname{C}(\Delta_XT),\quad \text{C denotes contraction,}\label{Delta-contraction}
\end{align}
\end{subequations}
where $f$ is a function and $T,U$ are general tensors.

Dependence on the generating vector field is more subtle.  The map
$X\mapsto\Delta_X$ is linear over constants, but not over functions.  From
\begin{align}
 (A_{fX})_{\mu\nu}
 =f(A_X)_{\mu\nu}+(\nabla_{[\mu}f)X_{\nu]},
\end{align}
one obtains the representation-independent formula
\begin{align}
 \Delta_{fX}T
 =f\Delta_XT+\r(B_{f,X})T,
 \qquad
 (B_{f,X})_{\mu\nu}
 \equiv(\nabla_{[\mu}f)X_{\nu]}.
 \label{Delta-fX-general}
\end{align}
For a vector $V$, this reads
\begin{align}
 (\Delta_{fX}V)^\mu
 =f(\Delta_XV)^\mu
 +\frac12\left[(X\cdot V)\nabla^\mu f-V[f]X^\mu\right].
 \label{Delta-fX-vector}
\end{align}
With the generator convention written as $\r(B)=-\tfrac{\i}{2}B_{\mu\nu}\Si^{\mu\nu}$, \eqref{Delta-fX-general} is equivalently
\begin{align}
 \Delta_{fX}T
 =f\Delta_XT
 -\frac{\i}{4}(\d f\wedge X^\flat)_{\mu\nu}
 \Si^{\mu\nu}T.
 \label{Delta-fX-generators}
\end{align}

These identities delimit the transport interpretation. Since
$\Delta_XT$ at a point depends on $A_X=\nabla_{[\mu}X_{\nu]}$, it depends
on the first jet of $X$, not only on the tangent vector $X$ at that point. Equivalently, \eqref{Delta-fX-general} shows that $\Delta$ is not an affine connection with $X$ as its first argument. Therefore,
\begin{align}
 \Delta_XT=0
\end{align}
is an ordinary differential equation once a vector field $X$ and its first derivatives have been specified, but it does not define parallel transport determined only by an unparametrized curve.  

There is nevertheless a precise pointwise geometric statement. The endomorphism $A_X$ is antisymmetric and hence belongs to
$\mathfrak{so}(TM,g)$.  Thus
\begin{align}
 \Delta_XT
 =\nabla_XT+\r(A_X)T
 =\nabla_XT-\frac{\i}{4}(\d X^\flat)_{\mu\nu}
 \Si^{\mu\nu}T
 \label{Delta-local-Lorentz-meaning}
\end{align}
is the covariant derivative along $X$ supplemented by the infinitesimal
orthogonal rotation determined by the antisymmetric part of $\nabla X$.

The global geometric answer is postponed to
section~\ref{sec:metric-lie-derivative}.  There, $\Delta_X$ is obtained
from a metric-dependent lift of the infinitesimal diffeomorphism to the
orthonormal frame bundle.  This explains why the first jet of $X$ enters
and why the construction is not an affine connection on spacetime.

\subsection{Commutators}\label{All comm}
Directly applying the action rule, we can compute all the independent commutators involving the covariant derivative, Lie derivative, and covariant variation:
\begin{subequations}\label{comm V}
\begin{align}
[\nabla_\r,\nabla_\s]V^\m&=R^\m{}_{\n\rs}V^\n,\\
[\cl_X,\cl_Y]V^\m&=\cl_{[X,Y]}V^\m,\\
[\Delta_X,\Delta_Y]V^\m&=\Delta_{[X,Y]}V^\m+(\ca_{X, Y})^\m{}_\n V^\n\label{eq3.32c},\\
[\cl_X,\nabla_\n]V^\m&=\cl_X\Gamma_{\n\r}^\m V^\r,\label{3.76d}\\
[\Delta_X,\nabla_\n]V^\mu&=[\Delta_X,\p_\n]V^\mu+\Delta_X\Gamma_{\n\r}^\m V^\r,\label{3.76e}\\
[\mathcal{L}_X,\Delta_Y]V^\mu&=\mathcal{L}_{[X,Y]} V^\m+\cl_X(S_Y)^\m_\n V^\n,\label{LXDelY}
\end{align}
\end{subequations}
where to compute \eqref{eq3.32c}, we need an identity
\begin{align}
\mathcal{L}_X (S_Y)^\mu_\nu - \mathcal{L}_Y (S_X)^\mu_\nu = (S_{[X,Y]})^\mu_\nu - 2 [S_X, S_Y]^\mu{}_\nu.
\end{align}
Moreover, \eq{3.76d} defines the tensorial Lie variation of the connection, cf., appendix~\ref{Lie conn}. As a comparison,~\eqref{3.76e} fixes only the sum of the partial-derivative commutator and a chosen connection variation. Their separate definitions are discussed in appendix~\ref{CVch}.

Similarly, we obtain the results for a one-form
\begin{subequations}
\begin{align}
[\nabla_\r,\nabla_\s]A_\m&=-R^\n{}_{\m\rs}A_\n,\\
[\cl_X,\cl_Y]A_\m&=\cl_{[X,Y]}A_\m,\\
[\Delta_X,\Delta_Y]A_\m&=\Delta_{[X,Y]}A_\m+(\ca_{X, Y})_\m{}^\n A_\n,\\
[\cl_X,\nabla_\n]A_\m&=-\cl_X\Gamma_{\n\m}^\r A_\r,\\
[\Delta_X,\nabla_\n]A_\mu&=[\Delta_X,\p_\n]A_\mu-\Delta_X\Gamma_{\n\m}^\r A_\r,\\
[\mathcal{L}_X,\Delta_Y]A_\mu&=\mathcal{L}_{[X,Y]} A_\m -\cl_X(S_Y)_\m^\n A_\n.
\end{align}
\end{subequations}

The above are the component formulas in fixed coordinates. In particular, the derivative labels are held fixed, and $\mathcal L_X\Gamma^\rho_{\mu\nu}$ denotes the tensorial Lie variation of the connection coefficients, cf., appendix \ref{Lie conn}. The coordinate-free formulas below instead transport the differentiating vector field and use the geometric commutator. These two conventions must not be mixed.

\paragraph{A coordinate-free formulation.}
Equations in \eq{comm V} use components. A coordinate-free formulation is more concise
\begin{subequations}
\begin{align}
[\nabla_X,\nabla_Y]V
&=\nabla_{[X,Y]}V+R(X,Y)V,\\
[\cl_X,\cl_Y]V&=\cl_{[X,Y]}V,\\
[\Delta_X,\Delta_Y]V&=\Delta_{[X,Y]}V+\r(\ca_{X, Y})V,\\
[\cl_X,\nabla_Y]V&=\nabla_{[X,Y]}V+P_X(Y,V),\label{aff Lie def}\\
[\Delta_X,\nabla_Y]V&=\nabla_{[X,Y]} V+P_X(Y,V)-\r(\nabla_YS_X)V,\label{aff cov def}\\
[\mathcal{L}_X,\Delta_Y]V&=\mathcal{L}_{[X,Y]} V+\r(\cl_XS_Y)V.
\end{align}
\end{subequations}
Here we introduce $P_X\equiv\cl_X\nabla$, or more explicitly
\begin{align}
P_X(Y,V)= (\cl_X\nabla)_YV=\cl_X(\nabla_YV)
 -\nabla_{\cl_XY}V-\nabla_Y(\cl_XV)
\end{align}
in \eq{aff Lie def}, which agrees with the definition of the Lie derivative of the affine connection
\begin{align}
\cl_X\nabla=\frac{\d}{\d t}\bigg|_{t=0}\phi^*_t\nabla,
\end{align}
where $\phi_t$ is the flow generated by $X$. In \eq{aff cov def}, we can similarly treat the last two terms as 
\begin{align}
Q_X\equiv \De_X\nabla=P_X-\nabla S_X.
\end{align}
Here $Q_X$ is fixed by the operator equation, and should not be identified automatically with a separately chosen fixed-coordinate coefficient variation $\Delta_X\Gamma^\rho_{\mu\nu}$. Appendix \ref{CVch} keeps these two notions apart. Moreover, the Jacobi identities concerning these operators are discussed in appendix \ref{sec jacobi}.

\subsection{Action on differential forms}
In this subsection, we discuss how the covariant variation acts on the differential forms. In particular, we will explore the relation between the covariant variation and the existing operators, including the exterior differential, interior product and Hodge duality.

A $p$-form $\a_{\m_1\cdots\m_p}$ is a totally antisymmetric covariant tensor, and we can directly apply the slotwise rule to obtain
\begin{subequations}
\begin{align}
\De_X\a_{\m_1\cdots\m_p}
&=\cl_X\a_{\m_1\cdots\m_p}-p(S_X)_{[\m_1}^\n \a_{|\n|\m_2\cdots\m_p]}\\
&=\nabla_X\a_{\m_1\cdots\m_p}+p(A_X)_{[\m_1}{}^\n \a_{|\n|\m_2\cdots\m_p]}.
\end{align}
\end{subequations}
For the wedge product, we find
\begin{align}
\Delta_X(\a\wedge\b)=\Delta_X\a\wedge\b+\a\wedge\De_X\b,
\end{align}
just like the Lie derivative. 

Recall the Cartan formula
\begin{align}
\cl_X=i_X\d+\d i_X,
\end{align}
from which we know 
\begin{align}
[\cl_X,\d]=0 \qaq [\cl_X,i_X]=0.
\end{align}
Then it is easy to find
\begin{align}
[\Delta_X,\d]&=[\r(S_X),\d],\\
[\Delta_X,i_X]&=[\r(S_X),i_X].
\end{align}
More explicitly, we have
\begin{align}
([\Delta_X,\d]\a)_{\m_0\m_1\cdots\m_p}&=-(p+1)(S_X)_{[\m_0}^\n\nabla_{|\n|}\a_{\m_1\cdots\m_p]} + p(p+1)\nabla_{[\m_0}(S_X)_{\m_1}^\n\a_{|\n|\m_2\cdots\m_p]}.
\end{align}

Instead of $[\cl_X,i_X]=0$, a more useful result is
\begin{align}
[\cl_X,i_Y]=i_{\cl_XY}=i_{[X,Y]}.
\end{align}
For the covariant variation, a similar relation holds
\begin{align}
[\Delta_X,i_Y]=i_{\Delta_XY}.
\end{align}

It is known that the useful forms with higher rank (e.g., co-dimension 0, 1, and 2) can be related to the lower rank one through Hodge duality. We will discuss in detail how the covariant variation acts on them in appendix \ref{higher forms}.

Now we involve Hodge duality. There are some identities
\begin{subequations}
\begin{align}
\d*\a&=\nabla_{\m_p}\a^{\m_1\cdots\m_p}(\d^{d-p+1}x)_{\m_1\cdots\m_{p-1}}\qquad 1\le p\le d,\label{stokesform}\\
i_X*\a&=(-1)^p(p+1)X^{[\m}\a^{\m_1\cdots\m_p]}(\d^{d-p-1}x)_{\m\m_1\cdots\m_p}\qquad 0\le p\le d-1,\\
\cl_X*\a&=\bigl[(-1)^{p-1}pX^{[\m_1}\nabla_{\m}\a^{\m_2\cdots\m_p]\m}+(p+1)\nabla_{\m}(X^{[\m}\a^{\m_1\cdots\m_p]})\bigr](\d^{d-p}x)_{\m_1\cdots\m_p}.
\end{align}
\end{subequations}
Moreover, the Lie derivative does not commute with the Hodge duality
\begin{align}
[\mathcal{L}_{X}, *] \a &= (\nabla \cdot X) * \a - 2p (S_X)^{\mu_{1}}_\n \a^{\n\mu_{2} \cdots \mu_{p}} (\d^{d-p} x)_{\m_{1} \cdots \mu_{p}}.
\end{align}

The Hodge star is best treated without expanding it in a degree-mismatched basis. We know
\begin{align}
[\r(S_X)\a]_{\m_1\cdots\m_p}
=-\sum_{i=1}^p(S_X)_{\m_i}^\n
\a_{\m_1\cdots\m_{i-1}\n\m_{i+1}\cdots\m_p}=-p(S_X)_{[\m_1}^\n
\a_{|\n|\m_2\cdots\m_p]},
\end{align}
then
\begin{align}
[\cl_X,*]\a=(\nabla\mathbin{\cdot}X)*\a+2*[\r(S_X)\a].
\end{align}
Moreover, because $\Delta_Xg=0$ and $\Delta_X\bm\ep=0$, the correction by $S_X$ cancels this metric variation, and hence
\begin{align}
\Delta_X(*\a)=*(\Delta_X\a)=(\Delta_X\a)^{\m_1\cdots\m_p}(\d^{d-p}x)_{\m_1\cdots\m_p}.
\end{align}

\section{Extended differential operators}\label{sec:related-operators}
Several related constructions appeared in the literature, including the spinor Lie derivative \cite{Lichnerowicz1963,Kosmann:1971ugf,Penrose:1986ca,HABERMANN1996131,Godina:2003tc,Godina:2005mt,Leao:2015qqc,Helfer:2016zvl}, Lie--Lorentz derivative \cite{Ortin:2002qb,Fatibene:2009,Elgood:2020svt}, Lorentz-Lie derivative \cite{Jacobson:2015uqa}, and Kosmann derivative \cite{Aneesh:2020fcr}. We will adopt the last name in this paper.  They all supplement the infinitesimal diffeomorphism generated by $X$ with a metric-dependent local Lorentz rotation. However, the operators act on different bundles, so they should not be identified before their domains and representations are specified. The generalized Lie derivative of double field theory acts instead on a generalized tangent bundle and is a distinct construction \cite{Grana:2008yw,Hohm:2010pp,Angus:2018mep,Ballesteros:2026irl}.

In this section, we will first review the Kosmann derivative, and then propose a total covariant variation from combining the Kosmann derivative and covariant variation. These two can be unified to the metric Lie derivative. In the last subsection, we explore two families of operators which contain all three spacetime operators, i.e., affine connection, Lie derivative, and covariant variation.

\subsection{Kosmann derivative}
Let $e^a{}_{\mu}$ be a vielbein, with
\begin{align}
g_{\mu\nu}=\eta_{ab}e^a{}_{\mu}e^b{}_{\nu},
 \qquad
 \nabla_\mu e^a{}_{\nu}
 =\partial_\mu e^a{}_{\nu}-\Gamma^\rho_{\mu\nu}e^a{}_{\rho}
 +\omega_\mu{}^a{}_b e^b{}_{\nu}=0.
\end{align}
For a vector field $X$, define the antisymmetric Lorentz parameter
\begin{align}
 (\lambda_X)^{ab}
 &\equiv e^{\mu[a}\mathcal{L}_X e^{b]}{}_{\mu}
 =X^\mu\omega_\mu{}^{ab}+(A_X)^{ab}.
\end{align}
Let $E_\varrho$ be a bundle associated to the orthonormal frame bundle, or to a chosen spin group lift when spinors are present, and let $\r_{\mathrm L}$ denote the induced Lorentz-algebra action. For a section $\Phi\in\Gamma(E_\varrho)$, the Kosmann derivative is
\begin{align}
 \mathcal{K}_X\Phi
 &\equiv\mathcal{L}_X\Phi+\r_{\mathrm L}(\lambda_X)\Phi
 =\nabla_X\Phi+\r_{\mathrm L}(A_X)\Phi.
 \label{KosmannD}
\end{align}
In the first expression, $\mathcal{L}_X$ differentiates the Lorentz components as spacetime scalars. The parameter $\lambda_X$ is not itself Lorentz covariant whose inhomogeneous transformation cancels that of $\mathcal{L}_X\Phi$. The second expression is manifestly Lorentz covariant.

With the following convention for Lorentz generator
\begin{align}
 (\Si^{cd})^a{}_b
 &=\i\bigl(\eta^{ca}\delta^d_b-\eta^{da}\delta^c_b\bigr),
 \qquad
 \r_{\mathrm L}(A_X)=-\frac{\i}{2}(A_X)_{cd}\Si^{cd},
 \label{LorentzG}
\end{align}
the Kosmann derivative \eqref{KosmannD} gives, for a Lorentz tensor of type $(1,1)$,
\begin{align}
 \mathcal{K}_Xt^a{}_b
 &=X^\mu\nabla_\mu t^a{}_b
 +(A_X)^a{}_c t^c{}_b+(A_X)_b{}^c t^a{}_c
 =\nabla_Xt^a{}_b-\frac{\i}{2}(A_X)_{cd}(\Si^{cd}t)^a{}_b.
 \label{kosm}
\end{align}
For a spinor, set $\gamma^{ab}\equiv\gamma^{[a}\gamma^{b]}$ and the generator is $\Sigma^{ab}=\frac{\i}{2}\g^{ab}$. Then the spin-1/2 representation of the same formula is
\begin{align}
\mathcal{K}_X\psi =X^\mu\nabla_\mu\psi+\frac{1}{4}(A_X)_{ab}\gamma^{ab}\psi.
\end{align}
This is the classical Kosmann formula \cite{Kosmann:1971ugf}. It also agrees with equation (11) of \cite{Ortin:2002qb}. It is defined for an arbitrary vector field $X$, provided the relevant orthonormal or spin structure has been chosen. For a Killing field, the flow preserves $g$, and $\mathcal{K}_X$ is the infinitesimal action of that isometry on Lorentz tensors or spinors.

For generic vector fields, the Kosmann lift is not a homomorphism of Lie algebras. On any associated Lorentz bundle,
\begin{align}
 \bigl([\mathcal{K}_X,\mathcal{K}_Y]-\mathcal{K}_{[X,Y]}\bigr)\Phi
 =\r_{\mathrm L}\bigl(-[S_X,S_Y]\bigr)\Phi.
 \label{Kosmann-anomaly}
\end{align}
This is the representation-theoretic form of the curvature of the metric-space lift, cf., Proposition 4.10 and Remark 4.11 in \cite{Dalak:2026zbz}. The obstruction vanishes on the isometry algebra. More generally, it vanishes if either $X$ or $Y$ is conformal Killing, since then one of $S_X$ and $S_Y$ is proportional to the identity. 

We will discuss how the Kosmann derivative acts on the spin connection, and how it is linked with the covariant variation through the vielbein in appendices \ref{Kosm as cov Lie} and \ref{gra as gauge conn}.

\subsection{Total covariant variation}\label{sec total}
Now let $\Phi$ carry both spacetime indices and Lorentz tensor or spinor indices, so that it is a section of $T^k{}_lM\otimes E_\varrho$. As usual, $\nabla$ contains the Levi-Civita connection on the spacetime indices and the spin connection on the Lorentz indices. Denote their representation actions by $\r$ and $\r_{\mathrm L}$ as before, respectively. Combining both considerations of the covariant variation and Kosmann derivative, we can define
a total covariant variation
\begin{align}
 \pl_X\Phi
 \equiv\nabla_X\Phi
 +\r(A_X)\Phi+\r_{\mathrm L}(A_X)\Phi.
 \label{total-covariant-variation}
\end{align}
Equivalently, if $\mathcal{K}_X$ acts by the ordinary Lie derivative on the spacetime indices and by the Kosmann derivative on the Lorentz indices, while $\Delta_X$ acts on the spacetime indices and treats the Lorentz components as scalars, then
\begin{align}
 \pl_X
 =\mathcal{K}_X+\r(S_X)
 =\Delta_X+\r_{\mathrm L}(\lambda_X).
 \label{total-relations}
\end{align}
Consequently,
\begin{align}
 &\pl_Xg_{\mu\nu}=0,
 \qquad \pl_X\eta_{ab}=0,
 \qquad \pl_Xe^a{}_{\mu}=0,
 \qquad \pl_Xe_a{}^\mu=0,\\
 &\pl_XT^\mu{}_\nu=\Delta_XT^\mu{}_\nu,
 \qquad
 \pl_Xt^a{}_b=\mathcal{K}_Xt^a{}_b,
 \qquad
 \pl_X\psi=\mathcal{K}_X\psi.
\end{align}
In particular,
\begin{align}
 \pl_Xe^a{}_{\mu}
 &=\mathcal{K}_Xe^a{}_{\mu}-(S_X)_\mu{}^\nu e^a{}_{\nu}
 =\Delta_Xe^a{}_{\mu}+(\lambda_X)^a{}_b e^b{}_{\mu}=0,
\end{align}
where $\mathcal{K}_Xe^a{}_{\mu}=(S_X)_\mu{}^\nu e^a{}_{\nu}$. This identity is the precise sense in which the total operator is compatible with the soldering map between spacetime and Lorentz indices.

For example, the Kosmann derivative of a Rarita-Schwinger field (spin 3/2 field) is
\begin{align}
 \mathcal{K}_X\psi_\mu
 =X^\nu\nabla_\nu\psi_\mu+\psi_\nu\nabla_\mu X^\nu
 +\frac{1}{4}(A_X)_{ab}\gamma^{ab}\psi_\mu,
\end{align}
whereas its total covariant variation is
\begin{align}
 \pl_X\psi_\mu
 &=\mathcal{K}_X\psi_\mu-(S_X)_\mu{}^\nu\psi_\nu\\
 &=X^\nu\nabla_\nu\psi_\mu+(A_X)_\mu{}^\nu\psi_\nu
 +\frac{1}{4}(A_X)_{ab}\gamma^{ab}\psi_\mu.
\end{align}
The two operators therefore agree on a pure Lorentz field, but not on a generic mixed-index field unless the symmetric spacetime-index action vanishes. The total operator inherits the same obstruction to closure
\begin{align}
 \bigl([\pl_X,\pl_Y]-\pl_{[X,Y]}\bigr)\Phi
 =\bigl(\r+\r_{\mathrm L}\bigr)\bigl(-[S_X,S_Y]\bigr)\Phi.
\end{align}

\subsection{Metric Lie derivative}
\label{sec:metric-lie-derivative}
The metric Lie derivative supplies the geometric construction behind the covariant variation and the Kosmann derivative. The point is especially important for spinors: the spinor bundles associated with two different metrics are different bundles, so the ordinary difference quotient of their sections is not defined. A connection over the space of metrics gives the required identification. The original construction is due to Bourguignon and Gauduchon \cite{Bourguignon:1992sp}, while a recent generalization is given in \cite{Dalak:2026zbz}. 

For a vector space $V$, let $\operatorname{Met}_{r,s}(V)$ be the space of non-degenerate metrics of signature $(r,s)$. Along a curve $\gamma(t)\in\operatorname{Met}_{r,s}(V)$, define $\tau(t)\in\operatorname{End}(V)$ by
\begin{align}
 \tau(0)=\operatorname{id}_V,
 \qquad
 \dot\tau=-\frac{1}{2}\gamma^{-1}\dot\gamma\,\tau.
 \label{metric-space-transport}
\end{align}
Then $\tau(t)$ is an isometry from $(V,\gamma(0))$ to $(V,\gamma(t))$. Hence \eqref{metric-space-transport} defines a natural connection on the bundle of orthonormal frames over the space of metrics.

Let $\phi_t$ be the flow of $X$. Pushing a $g$-orthonormal frame forward by $\phi_t$ produces a frame orthonormal with respect to the transported metric. Parallel transport by \eqref{metric-space-transport} identifies it with a $g$-orthonormal frame and defines an ${\rm O}(r,s)$-invariant lift $X^{\mathrm O}$ of $X$ to the orthonormal frame bundle. The metric Lie derivative of an associated field $\Phi$ is
\begin{align}
 \mathcal{L}^g_X\Phi\equiv X^{\mathrm O}\Phi.
\end{align}
Theorem 2.7 of \cite{Dalak:2026zbz} gives, on a vector field $V$,
\begin{align}
 \mathcal{L}^g_XV
 =\mathcal{L}_XV+\frac{1}{2}g^{-1}(\mathcal{L}_Xg)V
 =\Delta_XV.
\end{align}
Because $\mathcal{L}^g_X$ obeys the Leibniz rule, commutes with contractions, and preserves $g$, this equality extends to every ordinary tensor bundle. Lifting $X^{\mathrm O}$ through the pin or spin double cover gives the Kosmann derivative, and using the induced action on mixed associated bundles gives the total covariant variation. Thus,
\begin{subequations}
\begin{align}
 \left.\mathcal{L}^g_X\right|_{\text{ordinary tensors}}
 &=\Delta_X,\\
 \left.\mathcal{L}^g_X\right|_{\text{Lorentz tensors or spinors}}
 &=\mathcal{K}_X,\\
 \left.\mathcal{L}^g_X\right|_{\text{mixed fields}}
 &=\pl_X.
 \label{metric-lie-relations}
\end{align}
\end{subequations}
These are three representations of the same metric-dependent frame lift, rather than three unrestrictedly identical operators. In particular, on an ordinary tensor $\mathcal{L}^g_X=\mathcal{L}_X$ only when the $S_X$ action vanishes, most notably when $X$ is Killing.

The connection \eqref{metric-space-transport} is not flat. Its curvature gives precisely the non-closure term in \eqref{Kosmann-anomaly}. Therefore, the metric Lie derivative is natural with respect to isomorphisms of the pair $(M,g)$, but it is not a representation of the full Lie algebra of vector fields at fixed $g$. Finally, the construction applies directly to fields associated to the orthonormal, pin, or spin frame bundle. It does not by itself define the variation of an affine connection or a spin connection, since connections are not tensor. Those cases require a separate affine or gauge-covariant definition.

\subsection{Families of tensor derivations}\label{sec fam cov}
The covariant derivative, ordinary Lie derivative, and covariant variation can be placed in a single two-parameter family
\begin{subequations}\label{two-parameter-family}
\begin{align}
D^{(\a,\b)}_X&=\nabla_X+\rho\bigl(\alpha A_X-\beta S_X\bigr)\\
&=(1-\a)\nabla_X+ \b\cl_X+(\a-\b)\Delta_X,
\end{align}
\end{subequations}
which reduces to the affine connection for $(0,0)$, the Lie derivative for $(1,1)$, and the covariant variation for $(1,0)$
\begin{align}
D^{(0,0)}_X=\nabla_X, \qquad D^{(1,0)}_X=\Delta_X, \qquad D^{(1,1)}_X=\mathcal L_X. \label{family-special-three}
\end{align}
The diagonal $\alpha=\beta=t$ is simply $D_X^{(t,t)}=(1-t)\nabla_X+t\mathcal L_X$, but the covariant variation lies off this diagonal. To be more explicit, we write its action on a vector or a one-form
\begin{subequations}
\begin{align}
D^{(\a,\b)}_X V^\m&=\nabla_X V^\m+\a \nabla^{[\m}X^{\n]}V_\n-\b \nabla^{(\m}X^{\n)}V_\n,\\
D^{(\a,\b)}_X A_\m&=\nabla_X A_\m+\a \nabla_{[\m}X_{\n]}A^\n+\b \nabla_{(\m}X_{\n)}A^\n.
\end{align}
\end{subequations}

For every $(\alpha,\beta)$, the operator acts on scalars as $X$, obeys the Leibniz rule, and commutes with contractions. It is therefore a first-order derivation of the tensor algebra. It is not an affine connection unless $(\alpha,\beta)=(0,0)$, because $A_{fX}$ and $S_{fX}$ contain derivatives of $f$. Its compatibility with the metric is measured by
\begin{align}
 D_X^{(\alpha,\beta)}g
 =2\beta S_X=\beta\mathcal L_Xg.
 \label{family-metric-compatibility}
\end{align}
Consequently, the entire line $\beta=0$ preserves the metric. Only this line acts by orthogonal generators and therefore lifts without further choices to the orthonormal, pin, or spin frame bundle. The point $(1,0)$ is distinguished on this line because it comes from the lift $X^{\rm O}$ described in the preceding subsection.

\paragraph{Curvature of the two-parameter family.}
Define the endomorphism-valued non-closure by
\begin{align}
\ca_{X,Y}^{(\alpha,\beta)}=[D_X^{(\alpha,\beta)},D_Y^{(\alpha,\beta)}] -D_{[X,Y]}^{(\alpha,\beta)}. \label{family-curvature-definition}
\end{align}
Before substituting special identities, direct expansion gives
\begin{align}
\ca_{X,Y}^{(\alpha,\beta)}
 &=R(X,Y)
 +\alpha\bigl(\nabla_XA_Y-\nabla_YA_X-A_{[X,Y]}\bigr)\nonumber\\
 &\quad
 -\beta\bigl(\nabla_XS_Y-\nabla_YS_X-S_{[X,Y]}\bigr)
 +[\alpha A_X-\beta S_X,\alpha A_Y-\beta S_Y].
 \label{family-curvature-first-expansion}
\end{align}
The closure of the ordinary Lie derivative and the known curvature of $\nabla_X$ imply
\begin{subequations}\label{AS-differential-identities}
\begin{align}
 \nabla_XS_Y-\nabla_YS_X-S_{[X,Y]}
 &=-[A_X,S_Y]-[S_X,A_Y],\\
 \nabla_XA_Y-\nabla_YA_X-A_{[X,Y]}
 &=-R(X,Y)-[A_X,A_Y]-[S_X,S_Y].
\end{align}
\end{subequations}
Substitution into \eqref{family-curvature-first-expansion} yields
\begin{align}
\ca_{X,Y}^{(\alpha,\beta)} &=(1-\alpha)R(X,Y) -\alpha(1-\alpha)[A_X,A_Y] +(\beta^2-\alpha)[S_X,S_Y]\nonumber\\
&\quad+\beta(1-\alpha) \bigl([A_X,S_Y]+[S_X,A_Y]\bigr).
\label{family-curvature-final}
\end{align}
This formula contains the familiar cases
\begin{align}
\ca^{(0,0)}_{X,Y}=R(X,Y), \qquad
\ca^{(1,0)}_{X,Y}=-[S_X,S_Y], \qquad
\ca^{(1,1)}_{X,Y}=0.\label{special anomaly}
\end{align}

There is a second flat point within this ordinary-tensor family:
\begin{align}
\ca^{(1,-1)}_{X,Y}=0.
\end{align}
It is not metric-compatible and does not define an orthonormal-frame or spin-frame lift, since $\beta=-1$. On vectors and covectors,
\begin{align}
D_X^{(1,-1)}V&=\bigl(\mathcal L_X(V^\flat)\bigr)^\sharp,\qquad D_X^{(1,-1)}A=\bigl(\mathcal L_X(A^\sharp)\bigr)^\flat.\label{conjugate-lie-derivative}
\end{align}
If $J_g$ denotes the multiplicative musical isomorphism that lowers every upper index and raises every lower index, then on any ordinary tensor bundle
\begin{align}
 D_X^{(1,-1)}=J_g^{-1}\mathcal L_XJ_g.
\end{align}
Its closure is therefore the closure of the ordinary Lie derivative in a metric-conjugate tensor realization. This construction does not define a Kosmann-type action on spinors.

Conversely, let $d\geq2$, let $\alpha,\beta$ be constants, and let $\rho$ be the faithful natural action on the full tensor algebra. Requiring $\ca^{(\alpha,\beta)}_{X,Y}=0$ for every pseudo-Riemannian metric of the fixed signature and all vector fields forces
\begin{align}
 \alpha=1,
 \qquad
 \beta=\pm1.
 \label{universally-flat-family-points}
\end{align}
Indeed, choosing first jets with $A_X=S_X=A_Y=S_Y=0$ at a point where $R(X,Y)\neq0$, equation \eqref{family-curvature-final} first gives $\alpha=1$. On a flat metric, choosing $A_X=A_Y=0$ and noncommuting symmetric first jets
$S_X,S_Y$, the same equation then gives $\beta^2=1$. In $d=1$, the curvature
and all endomorphism commutators vanish, so every parameter pair is flat.
Additional flat cases may occur on a restricted tensor representation, a special
background, or a restricted vector-field subalgebra.

Finally, the Jacobi identity gives
\begin{align}
\sum_{\text{cyclic}} \left( D^{(\alpha,\beta)}_X \mathcal{A}^{(\alpha,\beta)}_{Y,Z} + \mathcal{A}^{(\alpha,\beta)}_{X, [Y,Z]} \right) = 0.\label{family-bianchi-clean}
\end{align}
Equation \eqref{family-bianchi-clean}, rather than an ordinary scalar cocycle condition, is the appropriate identity when the family has nonzero curvature.

\paragraph{A three-parameter family.}
The irreducible decomposition of the covariant gradient is
\begin{align}
\nabla_\m X_\n=\nabla_{[\m}X_{\n]} + \nabla_{\langle\m}X_{\n\rangle}+\frac{1}{d}g_\mn \nabla\cdot X.
\end{align}
As before, the first term is denoted by $(A_X)_\mn$, while the remaining terms are denoted as
\begin{align}
(\wh S_X)_\mn&=\nabla_{\langle\m}X_{\n\rangle}=\nabla_{(\m}X_{\n)}-\frac{1}{d}g_\mn \nabla\cdot X,\\
(\phi_X)_\mn&=\frac{1}{d}g_\mn \nabla\cdot X\equiv\frac{1}{d}g_\mn\psi_X.
\end{align}
Now we define a three-parameter family
\begin{subequations}
\begin{align}
D^{(\a,\b,\g)}_X&=\nabla_X+\rho\bigl(\alpha \wh S_X+\beta \phi_X+\g A_X\bigr)\\
&=(1-\gamma)\nabla_X-\alpha\mathcal L_X+(\gamma+\alpha)\Delta_X+(\beta-\alpha)\rho(\phi_X).
\end{align}
\end{subequations}
The special choices give
\begin{align}
D_X^{(0,0,0)}=\nabla_X,\quad
D_X^{(0,0,1)}=\Delta_X,\quad
D_X^{(-1,-1,1)}=\mathcal L_X, \qaq D_X^{(1,1,1)}=J_g^{-1}\mathcal L_XJ_g.
\end{align}
The plane \(\alpha=\beta\) is the previous two-parameter family. Moreover, choosing $(0,-1,1)$ for a vector $\cy$ on $S^2$ gives
\begin{align}
D^{(0,1,1)}_\cy&=\nabla_\cy+\rho\bigl(-\phi_\cy+A_\cy\bigr)=\Delta_\cy-\rho(\phi_\cy),
\end{align}
which leads to the pure spherical action reduced from the bulk covariant variation in \eq{pure sph of bulk reduction}
\begin{align}
\wt\Delta_\cy\ca_A={\cal Y}^B\nabla_B\cA_A+\cA^B\nabla_{[A}{\cal Y}_{B]}+\frac{1}{2}\ca_A\nabla\cdot {\cal Y}.
\end{align}

This operator acts on the metric as
\begin{align}
D_X^{(\alpha,\beta,\gamma)}g
=-2\alpha\widehat S_X-2\beta\phi_X.
\end{align}
Therefore, it preserves the metric for any $X$ if and only if \(\alpha=\beta=0\), keeps the conformal structure for \(\alpha=0\), and preserves the volume form for \(\beta=0\), since
\begin{align}
D_X^{(\alpha,\beta,\gamma)}\bm\epsilon=-\beta\psi_X\bm\epsilon.
\end{align}

Now we define the three-parameter anomaly as
\begin{align}
\ca_{X,Y}^{(\alpha,\beta,\gamma)}=[D_X^{(\alpha,\beta,\gamma)},D^{(\alpha,\beta,\gamma)}_Y]-D^{(\alpha,\beta,\gamma)}_{[X,Y]}.
\end{align}
Since the trace term is closed
\begin{align}
\nabla_X\phi_Y-\nabla_Y\phi_X-\phi_{[X,Y]}=0,
\end{align}
the anomaly is independent of $\b$ and the result reads
\begin{align}
\ca^{(\alpha,\beta,\gamma)}_{X,Y}
&=(1-\gamma)R(X,Y)
-\gamma(1-\gamma)[A_X,A_Y]\nn\\
&\quad+(\alpha^2-\gamma)[\widehat S_X,\widehat S_Y]
-\alpha(1-\gamma)\bigl([A_X,\widehat S_Y]+[\widehat S_X,A_Y]\bigr).
\end{align}
It also satisfies a Bianchi-type identity
\begin{align}
\sum_{\mathrm{cyclic}}
\left(
D^{(\alpha,\beta,\gamma)}_X\ca^{(\alpha,\beta,\gamma)}_{Y,Z}
+\ca^{(\alpha,\beta,\gamma)}_{X,[Y,Z]}
\right)=0.
\end{align}

\section{Applications: Helicity of spinning fields}\label{sec:null-boundary-applications}
Let $M$ be a four-dimensional oriented Lorentz manifold with metric $g$, and let $\ch$ be a null hypersurface. Locally, or under an explicit global product assumption, write $\ch\simeq\R\times N$, where $N$ is a two-dimensional Riemann manifold with metric $q$. The pullback of the degenerate metric on $\ch$ to $N$ is $q$. The bulk Hodge duality maps the field strength $F$ to its duality $*_4F$, and commutes with the bulk covariant variation. Consider some bulk vector field $\xi_\cy\in TM$ parameterized by the vector $\cy\in TN$. As before, define the Levi-Civita tensor, volume form, induced metric variation and compensator on $N$
\begin{align}
\epsilon_{AB},\quad \mathrm d\Omega=\sqrt q\,\mathrm d^2x, \quad \cl_{\xi_\cy}g_{AB}\Big|_{\ch}=\de_\cy q_{AB}, \qaq S_{\mathcal Y}^q=\tfrac12q^{-1}\de_{\mathcal Y}q.
\end{align}
After pullback to tangent one-forms on $N$, the anomaly from covariant variation 
\begin{align}
\ca_{\cy,\cz}\ca_A=-\frac{1}{4}(\delta_{\cal Y}q_{AB}\delta_{\cal Z}q_{CD}-\delta_{\cal Z}q_{AB}\delta_{\cal Y}q_{CD})q^{BC}\cA^D
\end{align}
will always be proportional to the Hodge duality $*_2$ on $N$
\begin{align}
\ca_{\cy,\cz}\ca_A=-o_q(\cy,\cz) \ep_{AB}\ca^B,
\end{align}
with
\begin{align}
o_q(\mathcal Y,\mathcal Z)
&=\frac12\epsilon^{AB}
[S^q_{\mathcal Y},S^q_{\mathcal Z}]_{AB}.
\end{align}
Therefore, if this kind of bulk vector field changes the boundary metric nontrivially, we can expect the emergence of the helicity flux.

The theory in the Motivation is an example for this statement. In the rest of this section, we consider two extensions. The first one is to extend the story to a finite null hypersurface, and the second one is to consider general tensor fields.

\subsection{Electromagnetic field near a horizon}
In Gaussian null coordinates \((u,\rho,x^A)\), the near horizon geometry reads \cite{Tamburino:1966zz,Booth:2012xm,Donnay:2015abr,Donnay:2016ejv}
\begin{align}
\mathrm{d} s^2=-2\k\rho \mathrm{d} u^2-2\d u \mathrm{d} \rho+2\l_A\rho \mathrm{d} u\mathrm{d} x^A+(q_{AB}+\l_{AB}\rho)\mathrm{d} x^A\mathrm{d} x^B+O(\rho^2),\label{ds2}
\end{align}
where $\k,\l_A,q_{AB},\l_{AB}$ depend on $u,x^A$, while \(q_{AB}\) is non-degenerate and serves as the boundary metric. The surface $\ch$ at $\rho=0$ is the horizon of interest. For the canonical reduction below, we restrict to $\partial_uq_{AB}=0$. 

Preserving the gauge conditions and fall-offs leads to the following asymptotic Killing vector \cite{Donnay:2016ejv}
\begin{align}\label{xify}
\begin{aligned}
& \xi^u=f(u,\Omega), \\
& \xi^\rho=- \rho\partial_u f+\frac{1}{2} \rho^2\lambda^A \nabla_A f +O(\rho^3), \\
& \xi^A=\cY^A(\Omega)+  \rho\nabla^A f-\frac{1}{2} \rho^2\lambda^{AB} \nabla_B f +O(\rho^3),
\end{aligned}
\end{align}
where the indices of $\l^A,\l^{AB}$ are raised by $q^{AB}$, and $\nabla_A$ is adapted to $q_{AB}$. Its restriction at the horizon is
\begin{align}
\xi_{f,\cY}\Big|_{\ch}=f(u, \Omega)\p_u+\cY^A(\Omega)\p_A.
\end{align}

We are interested in exploring matter fields, for example the electromagnetic field $A_\m$, around the horizon. In particular, we ask whether the covariant variation matches the commutator between the flux generator and the field on the horizon, and what appears in the commutator between the fluxes.

Near the horizon, we impose \(A_\rho=0\) and require \(A_u\) to vanish on the horizon
\begin{align}
A_u=\sum_{k=1}^\infty \cA_u^{(k)}(u,\Omega)\rho^k,\qquad A_A=\sum_{k=0}^\infty \cA_A^{(k)}(u,\Omega)\rho^k.
\end{align}
For convenience, the label $(0)$ or $(1)$ for the leading order components will be neglected. We treat the leading order of \eq{ds2} as the background metric
\begin{align}
g_{uu}=g_{uA}=g_{\r\r}=g_{\r A}=0,\quad g_{u\rho}=-1,\qaq g_{AB}=q_{AB}.\label{emg}
\end{align}
The inverse background metric is 
\begin{align}
g^{uu}=g^{uA}=g^{\rho\rho}=g^{\rho A}=0,\quad g^{u\rho}=-1,\qaq
g^{AB}=q^{AB}.
\end{align}
Now we find
\begin{align}
A^u=0,\qquad A^A=\cA^A+O(\rho), \qquad A^\rho=\r\bigl(\l^A\cA_A-\cA_u\bigr)+O(\rho^2).
\end{align}
Computing the boundary symplectic form leads to
\begin{align}
\Omega(\delta \cA;\delta \cA)&=-\lim_{\rho\to 0}\int_{\ch_{\rho}} (\mathrm{d} ^{3}x)_\mu \delta F^{\mn}\wedge \delta A_\nu=\int \mathrm{d} u\mathrm{d}\Omega\,\delta {\cA}^A\wedge\delta \dot \cA_A,
\end{align}
where we have used \((\mathrm{d} ^{3}x)_\mu=-\delta_\mu^\rho \mathrm{d} u\mathrm{d}\Omega\). It gives rise to the fundamental commutator
\begin{align}
[\cA_A(u,\Omega),\dot \cA_B(u',\Omega')]=\frac{\i}{2}q_{AB}\delta(u-u')\delta(\Omega-\Omega').
\end{align}

\paragraph{Fluxes and their action.}
From the $\r$ component of equation of motion \(\nabla_\mu F^{\mu\nu}=0\), we get
\begin{align}
\dot{\cA}_u+\nabla_A\dot{\cA}^A=0 \qrq \cA_u=-\nabla_A\cA^A+\varphi(\Omega).\label{Au1}
\end{align}
Given this, the stress tensor
\begin{align}
T^{\mu{}}_{\ \ \nu}=F^{\mu\lambda}F_{\nu \lambda}-\frac{1}{4}\delta^\mu_\nu F^{}_{\lambda\rho}F^{\lambda\rho}
\end{align}
has the following relevant components
\begin{align}
T^\rho_{\ u}&=-\dot{\cA}^A\dot{\ca}_A,\qquad
T^\rho_{\ A}=-\dot{\ca}_A \nabla_B\cA^B-\dot{\ca}^B(\nabla_A\cA_B-\nabla_B\cA_A)+\vp\dot\ca_A.
\end{align}
at leading order. We can compute the flux associated with \eqref{xify}
\begin{subequations}
\begin{align} 
\mathcal{F}_f&=-\int \mathrm{d} u\mathrm{d}\Omega\, T^\rho_{\ u}\xi^u=\int \mathrm{d} u\mathrm{d}\Omega\, f(u,\Omega)\dot{\ca}^A\dot{\ca}_A,\\
\mathcal{F}_{\cY}&=-\int \mathrm{d} u\mathrm{d}\Omega\, T^\rho_{\ A}\xi^A=\int \mathrm{d} u\mathrm{d}\Omega\, \cY^A(\Omega)(\dot{\ca}^B\nabla^C\cA^D\bar P_{ABCD}-\vp\dot\ca_A),
\end{align}
\end{subequations}
where we have defined
\begin{align}
\bar P_{ABCD}=q_{AB}q_{CD}+q_{AC}q_{BD}-q_{AD}q_{BC}.
\end{align}
This $\bar P_{ABCD}$ has the same structure as $P_{ABCD}$, cf., \eq{Pabcd}. In the following, we will omit the soft part in $\cf_\cy$, since it is irrelevant to our purpose and can be discarded by choosing $\vp=0$.

We can compute the commutator between the fluxes and $\cA_A$
\begin{subequations}
\begin{align}
[\i\cf_f,\cA_A(u,\Omega)]&= f\dot{\ca}_A(u,\Omega),\label{FfA}\\
[\i\cf_\cY,\cA_A(u,\Omega)]&= \cY^B\nabla_B\cA_A+\frac{1}{2}\nabla\cdot \cY \cA_A+\cA^B\nabla_{[A}\cY_{B]}.\label{FYA}
\end{align}
\end{subequations}
These results can be compared with the corresponding Lie derivatives. From the bulk Lie derivatives of $A_A$ along $\xi_{f}$ and $\xi_{\cal Y}$, we find
\begin{align}
\cl_{\xi_f}A_A=f\dot{\ca}_A+O(\rho) \qrq \delta_f\cA_A=f\dot{\ca}_A,\label{Lxifa}
\end{align}
and
\begin{align}
\cl_{\xi_{\cal Y}}A_A&=\cY^B\p_B\cA_A+\cA_B\p_A\cY^B+O(\rho) \qrq \delta_\cY \cA_A=\cY^B\nabla_B\cA_A+\cA_B\nabla_A\cY^B.\label{LxiYa}
\end{align}
\eqref{Lxifa} agrees with \eqref{FfA}, but \eqref{LxiYa} does not agree with \eqref{FYA}. 

This is expected since we should modify the Lie derivative by the variation of the metric to get the covariant variation
\begin{align}
\Delta_{\xi_{f/\cY}} A_A=\cl_{\xi_{f/\cY}} A_A-\frac{1}{2}\cl_{\xi_{f/\cY}} g_{A\m}A^\m=\Delta_{f/\cY}\cA_A+O(\r).
\end{align}
More explicitly, one can compute
\begin{align}
&\cl_{\xi_f}g_{uA}=O(\r) \qaq \cl_{\xi_f}g_{AB}=O(\r),
\end{align}
and
\begin{align}
\cl_{\xi_{\cal Y}}g_{uA}=O(\r),\qquad \cl_{\xi_{\cal Y}}g_{AB}=\de_\cY q_{AB}+O(\r).
\end{align}
Thus only $\delta_\cY A_A$ will receive non-trivial correction and the result is
\begin{align}
\Delta_\cY \cA_A=\delta_\cY \cA_A-\frac{1}{2}\delta_\cY  q_{AB}\cA^B=\cY^B\nabla_B\cA_A+\cA^B\nabla_{[A}\cY_{B]},
\end{align}
which agrees with the commutator \eqref{FYA} for the divergence-free $\cY^A$. The divergence-free condition is expected, since otherwise the bulk vector
\begin{align}
\xi_\cy=\cy^A\p_A+O(\r)
\end{align}
is not divergence-free. On the other hand, one can verify that the superrotation vector field \eq{xiy} is divergence-free for any vector $\cy^A$ on the sphere.

\paragraph{Helicity flux.}
The extension from divergence-free $\cY^A$ to an arbitrary vector field requires care at a finite Gaussian-null horizon. For the asymptotic Killing vector \eqref{xify}, $\xi^\rho=-\rho\partial_u f+O(\rho^2)$, and therefore
\begin{align}
\left.\nabla_\mu\xi^\mu\right|_{\ch}=\partial_u f-\partial_u f+\nabla_A\cY^A=\nabla_A\cY^A.\label{hor-div}
\end{align}
Thus no choice of $f$, including $f=\tfrac12u\nabla_A\cY^A$, makes this bulk vector divergence-free when $\nabla_A\cY^A\neq0$.\footnote{\label{fn df bulk} In fact, we can construct a bulk vector
\begin{align}
\xi_\cy=-\frac12u\nabla\cdot\cy\p_u-\frac12\r\nabla\cdot\cy\p_\r+\cy^A\p_A,\nn
\end{align}
which has vanishing divergence for any smooth $\cy$. However, the action of the flux derived from this vector field still does not agree with the corresponding covariant variation, due to the absence of the divergence term in the boundary metric variation. Compared to the case of null infinity, the difference is that in the latter case, we need to rescale $\eta_{AB}=r^2\g_{AB}$ to obtain celestial metric $\g_{AB}$, while the order of finite horizon here is just $\r^0$, and no rescaling is needed.} Moreover,
\begin{align}
\left.\cl_\xi g_{AB}\right|_{\ch}=\cl_\cY q_{AB},\label{hor-metric-var}
\end{align}
not $\cl_{\cY}q_{AB}-(\nabla\mathbin{\cdot}\cY)q_{AB}$. Equations \eqref{hor-div} and \eqref{hor-metric-var} follow directly from the Gaussian null expansion and show that the divergence term familiar at null infinity cannot be imported to a finite horizon.

We therefore restrict this subsection to $\nabla_A\cY^A=0$. This condition is preserved by the Lie bracket. Defining
\begin{align}
\Th^{(q)}_{AB}(\cY)\equiv\cl_{\cY}q_{AB}=2\nabla_{(A}\cY_{B)},\qquad q^{AB}\Th^{(q)}_{AB}(\cY)=0,
\end{align}
the boundary covariant variation is
\begin{align}
\Delta_\cY\cA_A
=\delta_\cY\cA_A-\frac12\Th^{(q)}_{AB}(\cY)\cA^B
=\cY^B\nabla_B\cA_A+\cA^B\nabla_{[A}\cY_{B]},\label{hor-cov-var}
\end{align}
which agrees with \eqref{FYA} on this subalgebra. Its non-closure follows from \eqref{special anomaly}:
\begin{align}
\ca_{\cY,\cZ}\cA_A
&\equiv[\Delta_\cY,\Delta_\cZ]\cA_A-\Delta_{[\cY,\cZ]}\cA_A
=-o_q(\cY,\cZ)\ep_{AB}\cA^B,\\
o_q(\cY,\cZ)
&\equiv\frac14\ep^{AB}\Th^{(q)}_{AC}(\cY)\Th^{(q)C}{}_{B}(\cZ).\label{hor-anomaly}
\end{align}
The last equality uses only the fact that every two-form on a two-dimensional Riemannian manifold is proportional to its volume form. We can construct a flux
\begin{align}
\co_h=\int\d u\d\Omega\,h(\Omega)\ep_{AB}\cA^{A}\dot{\cA}^{B},
\end{align}
which acts as $[\i\co_h,\cA_A]=-h\ep_{AB}\cA^B$. Hence \eqref{hor-anomaly} identifies the possible non-central extension of the flux algebra with $\co_{o_q(\cY,\cZ)}$. 


\subsection{Tensor fields in arbitrary dimensions}
In \cite{Liu:2024rvz}, we have explored the electromagnetic helicity flux in higher dimensions. Beyond 4 dimensions, the story is not as perfect as before. It is not necessary to always start from a Lagrangian to construct the flux, and consider its action. Instead, we can perform the covariant variation of a general tensor field along a specific vector field near a null hypersurface (e.g., the superrotation $\xi_\cy$ near $\ci^+$), and expect the anomaly will tell us how to define the helicity rotation for this field.

Let $m=d-2>0$, and write the Minkowski metric near future null infinity as
\begin{align}
\d s^2=-\d u^2-2\d u\d r+r^2\gamma_{AB}\d x^A\d x^B,
\end{align}
where $\gamma_{AB}$ is the metric on $S^m$, and $\nabla_A$ denotes its Levi-Civita connection. For a smooth, $u$-independent vector field $\cY^A$ on $S^m$, the bulk extension needed below is
\begin{subequations}\label{general d superrotation}
\begin{align}
\xi_\cY^u&=\frac{u}{m}\nabla\cdot\cy,\\
\xi_\cY^r&=-\frac{r}{m}\nabla\cdot\cy+\frac{u}{m^2}\nabla^2\nabla\cdot\cy+O(r^{-1}),\\
\xi_\cY^A&=\cY^A-\frac{u}{mr}\nabla^A\nabla\cdot\cy+O(r^{-2}),
\end{align}
\end{subequations}
which is divergence free, $\nabla_\mu\xi_\cY^\mu=O(r^{-2})$. Its action on the rescaled angular metric is
\begin{align}
\delta_\cY\gamma_{AB}
&\equiv\lim_{r\to\infty}r^{-2}\cl_{\xi_\cY}g_{AB}
=\nabla_A\cY_B+\nabla_B\cY_A-\frac{2}{m}\gamma_{AB}\nabla\cdot\cy
\equiv\Th_{AB}(\cY),\label{general d theta}
\end{align}
and $\gamma^{AB}\Th_{AB}(\cY)=0$. For $m=2$, this reduces to \eqref{deygab} and to the angular part of \eqref{dyemn}. The arbitrary-dimensional Maxwell reduction of \eqref{general d superrotation} is given in \cite{Liu:2024rvz}.

Consider a covariant rank-$k$ tensor\footnote{This subsection is kinematical rather than a proposal for universal tensor-field dynamics. Once an irreducible Young type and a definite gauge realization are specified, the free field equation and Lagrangian are known in several formulations: the constrained Labastida theory \cite{Labastida:1987kw}, geometric second-order equations and actions that are generally nonlocal before gauge fixing or the introduction of auxiliary fields \cite{deMedeiros:2003osq}, nonlocal quadratic actions \cite{Bekaert:2006ix}, local unconstrained metric-like actions with compensators and Lagrange multipliers \cite{Campoleoni:2008jq}, and BRST-BFV actions for massless and massive bosonic fields with arbitrary Young tableaux on an enlarged auxiliary-field space \cite{Buchbinder:2011xw}. A general rank-$k$ tensor can contain several Young sectors, so its rank and kinematical transformation law alone do not determine its gauge constraints, equations of motion, or reduced symplectic structure. The discussion below therefore assumes only the stated fall-off, transversality, and a boundary symplectic structure.} with an arbitrary index symmetry. In an asymptotically orthonormal frame, assume the fall-off
\begin{align}
T_{\m_1\cdots\m_k}
=r^{-\Delta}\cT_{\m_1\cdots\m_k}(u,\Omega)
+O(r^{-\Delta-1}).
\end{align}
Passing to retarded coordinates contributes one factor of $r$ for each covariant angular index. Hence its purely angular component is
\begin{align}
T_{A_1\cdots A_k}
=r^{k-\Delta}\cT_{A_1\cdots A_k}(u,\Omega)
+O(r^{k-\Delta-1}).\label{tensor angular falloff}
\end{align}
The fall-off \eqref{tensor angular falloff} alone is not sufficient to obtain a closed transformation law for $\cT_{A_1\cdots A_k}$. In fact, if
\begin{align}
T_{A_1\cdots A_{i-1}rA_{i+1}\cdots A_k}
=r^{k-1-\Delta}\cT^{(r,i)}_{A_1\cdots\widehat{A_i}\cdots A_k}
+O(r^{k-2-\Delta}),
\end{align}
then direct evaluation of the Lie derivative gives
\begin{align}
\delta_\cY\cT_{A_1\cdots A_k}
&\equiv\lim_{r\to\infty}r^{\Delta-k}
\cl_{\xi_\cY}T_{A_1\cdots A_k}\nn\\
&=\left(
\frac{u}{m}\nabla\cdot\cy\partial_u
+\cl_\cY
+\frac{\Delta-k}{m}\nabla\cdot\cy
\right)\cT_{A_1\cdots A_k}\nn\\
&\quad
-\frac{1}{m}\sum_{i=1}^{k}
\nabla_{A_i}\nabla\cdot\cy\,
\cT^{(r,i)}_{A_1\cdots\widehat{A_i}\cdots A_k}.
\label{general tensor lie variation}
\end{align}
Here
\begin{align}
\cl_\cY\cT_{A_1\cdots A_k}
=\cY^C\nabla_C\cT_{A_1\cdots A_k}
+\sum_{i=1}^{k}\nabla_{A_i}\cY^B
\cT_{A_1\cdots A_{i-1}BA_{i+1}\cdots A_k}.
\end{align}
Thus the angular radiative field is closed only after imposing
\begin{align}
\cT^{(r,i)}_{A_1\cdots\widehat{A_i}\cdots A_k}=0,
\qquad i=1,\cdots,k,\label{radial projection condition}
\end{align}
or a stronger fall-off. In Cartesian coordinates, \eqref{radial projection condition} is the leading transversality condition $n^{\mu_i}\cT_{\mu_1\cdots\mu_k}=0$. For the gauge potential, it may follow from the radial gauge.

Under \eqref{radial projection condition}, the boundary reduction of the covariant variation
\begin{align}
\Delta_\cY\cT_{A_1\cdots A_k}
&=\delta_\cY\cT_{A_1\cdots A_k}
-\frac{1}{2}\sum_{i=1}^{k}\Th_{A_i}{}^B(\cY)
\cT_{A_1\cdots A_{i-1}BA_{i+1}\cdots A_k}\nn\\
&=\left(\frac{u}{m}\nabla\cdot\cy\partial_u+\cY^C\nabla_C+\frac{\Delta}{m}\nabla\cdot\cy\right)\cT_{A_1\cdots A_k}\nn\\
&\quad +\sum_{i=1}^{k}(A_\cy)_{A_i}{}^B
\cT_{A_1\cdots A_{i-1}BA_{i+1}\cdots A_k},
\label{general tensor cov variation}
\end{align}
where
\begin{align}
(A_\cy)_A{}^B\equiv\gamma^{BC}\nabla_{[A}\cY_{C]}.
\end{align}
The cancellation of the $k$-dependent trace terms in \eqref{general tensor cov variation} is important: the conformal weight is $\Delta/m$, whereas the spin is carried entirely by the antisymmetric endomorphism $A_\cy$. In particular, $\Delta_\cY\gamma_{AB}=0$, so traces and fixed Young symmetries are preserved.

Using \eqref{special anomaly} with $(S_\cY)_A{}^B=\tfrac12\Th_A{}^B(\cY)$, define the two-form
\begin{align}
o_{AB}(\cY,\cZ)
\equiv\frac{1}{4}\left[
\Th_{AC}(\cY)\Th^C{}_B(\cZ)
-\Th_{AC}(\cZ)\Th^C{}_B(\cY)
\right],
\qquad o_{AB}=-o_{BA}.\label{general d o}
\end{align}
The non-closure on the covariant tensor is then
\begin{align}
\ca_{\cY,\cZ}\cT_{A_1\cdots A_k}
&\equiv
[\Delta_\cY,\Delta_\cZ]\cT_{A_1\cdots A_k}
-\Delta_{[\cY,\cZ]}\cT_{A_1\cdots A_k}\nn\\
&=-\sum_{i=1}^{k}o_{A_i}{}^B(\cY,\cZ)
\cT_{A_1\cdots A_{i-1}BA_{i+1}\cdots A_k}.
\label{general tensor anomaly}
\end{align}
If either vector is conformal Killing, its $\Th_{AB}$ vanishes and so does the anomaly. For $m=1$, every two-form vanishes, and correspondingly, there is no helicity rotation concerning $\SO(1)$. For $m=2$,
\begin{align}
o_{AB}(\cY,\cZ)=o(\cY,\cZ)\epsilon_{AB},\qquad
o(\cY,\cZ)=\frac14\epsilon^{AB}
\Th_{AC}(\cY)\Th^C{}_B(\cZ),
\end{align}
and \eqref{general tensor anomaly} reproduces \eqref{DYDZ}. For $m>2$, $o_{AB}$ has $m(m-1)/2$ local components and acts through the appropriate tensor representation of $\mathfrak{so}(m)$. This is a local rotation of the little group.

To construct the fluxes concerning the above boundary transformations, a symplectic structure is needed. The simplest choice is\footnote{In the totally symmetric Fronsdal sector, the bulk action and the radiative symplectic form were used for deriving higher-spin fluxes in \cite{Liu:2023jnc}. The related equations of motion, boundary conditions, finite actions, and asymptotic charges at null infinity in arbitrary dimensions were studied in \cite{Campoleoni:2017qot,Campoleoni:2025bhn}. These results motivate, but do not derive, a universal phase space for arbitrary Young symmetry. For the following choice, the invariant inner product, trace projection, gauge reduction, endpoint conditions, and normalization are additional inputs.}
\begin{align}
\Omega_\k=\k\int\d u\d\Omega\,\delta\ct^{A_1\cdots A_k}\wedge\delta\dot{\ct}_{A_1\cdots A_k}.
\label{radiative-symplectic-form1}
\end{align}
Then preserving the symplectic structure could fix the fall-off
\begin{align}
\de_{\De_\cy}\Omega_\kappa=0 \quad\Longleftrightarrow\quad \Delta=\frac m2.   
\end{align}
If restricted to $\nabla\cdot\cy=0$, the symplectic structure is preserved for any $\Delta$. From Hamilton's equation, we can determine the flux\footnote{To obtain the flux, we need also to impose appropriate endpoint behaviour about $u$ and some other technical conditions.}
\begin{align}
\cF_\cY&=\kappa\int\mathrm d u\mathrm d\Omega\,\dot{\mathcal T}^{A_1\cdots A_k}\Delta_\cy\ct_{A_1\cdots A_k}
\end{align}
for the boundary transformation $\Delta_\cy$ induced by the bulk field $\xi_\cy$. 

The anomaly's action can be generalized to
\begin{align}
\de_h\ct_{A_1\cdots A_k}=-\sum_{i=1}^{k}h_{A_i}{}^B(\Omega)
\cT_{A_1\cdots A_{i-1}BA_{i+1}\cdots A_k},
\end{align}
where $h_{AB}$ is a two-form on the sphere. Through Hamilton's equation, this transformation also leads to a helicity flux
\begin{align}
\co_h=\kappa\int\mathrm du\,\mathrm d\Omega\,\dot{\mathcal T}^{A_1\cdots A_k}\de_h\ct_{A_1\cdots A_k}.
\end{align}

As a first check, take a radiative one-form in radial gauge with the free-field fall-off $\Delta=m/2$. Equation \eqref{general tensor cov variation} becomes
\begin{align}
\Delta_\cY\cA_A
=\frac{u}{m}\nabla\cdot\cy\dot{\cA}_A
+\cY^B\nabla_B\cA_A
+\frac12\nabla\cdot\cy\cA_A
+\cA^B\nabla_{[A}\cY_{B]}.\label{general d vector variation}
\end{align}
Moreover, the helicity flux reads
\begin{align}
\co_{h}
=\int\d u\d\Omega\,
h_{AB}(\Omega)\dot{\cA}^{B}\cA^{A},
\end{align}
which agrees with \cite{Liu:2024rvz}. Taking $d=4$, the story in the Motivation is recovered.

\subsection{$p$-form in $d=2p+2$}
A useful example for arbitrary tensors is a $p$-form ($p>0$) gauge potential $B_{\m_1\cdots\m_p}$ in $d=2p+2$ dimensions, for which $m=2p$. In radial gauge, assume the radiative fall-off $\Delta=m/2=p$, and denote the angular component by $\cB_{A_1\cdots A_p}$. Then
\begin{align}
\Delta_\cY\cB_{A_1\cdots A_p}
&=\left(
\frac{u}{2p}\nabla\cdot\cy\partial_u
+\cY^C\nabla_C
+\frac12\nabla\cdot\cy
\right)\cB_{A_1\cdots A_p}\nn\\
&\quad
+\sum_{i=1}^{p}(A_\cy)_{A_i}{}^B
\cB_{A_1\cdots A_{i-1}BA_{i+1}\cdots A_p}.
\end{align}
Equation \eqref{general tensor anomaly} is
\begin{align}
\ca_{\cY,\cZ}\cB_{A_1\cdots A_p}=-po_{[A_1}{}^B(\cY,\cZ)\cB_{|B|A_2\cdots A_p]}
\end{align}
In the following, we derive the helicity flux from this transformation.

The starting point is the free, non-chiral action
\begin{align}
S[B]&= -\frac{1}{2(p+1)!}\int\d^d x\sqrt{-g}\,H_{\mu_0\cdots\mu_p}H^{\mu_0\cdots\mu_p},
\label{p-form-free-action}
\end{align}
where $H_{\mu_0\cdots\mu_p}=(p+1)\nabla_{[\mu_0}B_{\mu_1\cdots\mu_p]}$. Its first variation is
\begin{align}
\delta S=\frac{1}{p!}\int\d^d x\sqrt{-g}\,(\nabla_{\mu_0}H^{\mu_0\mu_1\cdots\mu_p})\delta B_{\mu_1\cdots\mu_p}-\frac{1}{p!}\int_{\partial M}\d\Sigma_{\mu_0}\,H^{\mu_0\mu_1\cdots\mu_p}\delta B_{\mu_1\cdots\mu_p}.
\label{p-form-action-variation}
\end{align}
Therefore the covariant presymplectic current is
\begin{align}
\omega^\mu(\delta_1,\delta_2)=-\frac{1}{p!}\left(\delta_1H^{\mu\mu_1\cdots\mu_p}\delta_2B_{\mu_1\cdots\mu_p}-\delta_2H^{\mu\mu_1\cdots\mu_p}\delta_1B_{\mu_1\cdots\mu_p}\right).
\label{p-form-presymplectic-current}
\end{align}
In radial gauge, after solving the leading constraints and taking the finite radiative limit, the pullback of \eqref{p-form-presymplectic-current} to future null infinity gives
\begin{align}
\Omega_p=\frac{1}{p!}\int\d u\d\Omega\,\delta\cB^{A_1\cdots A_p}\wedge\delta\dot{\cB}_{A_1\cdots A_p}.
\label{p-form-radiative-symplectic-form}
\end{align}
The overall sign in \eqref{p-form-radiative-symplectic-form} fixes the null-boundary orientation and agrees with the Maxwell convention used above. 

The boundary transformation can be generally written as
\begin{align}
\delta_h\cB_{A_1\cdots A_p}=-ph_{[A_1}{}^B(\Omega)\cB_{|B|A_2\cdots A_p]},
\end{align}
where $h_A{}^B$ is a two-form on the sphere. Its Hamiltonian is
\begin{align}
\co_{h}^{(p)}=-\frac{1}{(p-1)!}\int\d u\d\Omega\,h_{A}{}^B\dot\cb^{AA_2\cdots A_p}\cB_{BA_2\cdots A_p}.
\label{p-form-invariant-generator}
\end{align}
Indeed, if the endpoint term at $u=\pm\infty$ vanishes, then
\begin{align}
\delta\co_{h}^{(p)}&=\frac{2}{(p-1)!}\int\d u\d\Omega\,h_{A}{}^B\delta\cB^{AA_2\cdots A_p}\dot\cB_{BA_2\cdots A_p}=i_{\delta_h}\Omega_p.
\label{p-form-hamiltonian-check}
\end{align}
In particular, we can verify
\begin{align}
[\i\co_{h}^{(p)},\cB_{A_1\cdots A_p}]=\de_h\cB_{A_1\cdots A_p}.
\label{p-form-generator-action}
\end{align}

\paragraph{Mode expansion.}
To exhibit the particle content, let $I=(A_1,\cdots,A_p)$ be a collective index and choose an orthonormal basis $E_I^\lambda(\Omega)$ such that
\begin{align}
\frac{1}{p!}E^{*\l I}E_I^\sigma=\delta^{\lambda\sigma},\qquad \lambda=1,\cdots,\binom{2p}{p}.
\label{p-form-polarization-normalization}
\end{align}
The nonzero-frequency radiative field has the expansion
\begin{align}
\cB_I(u,\Omega)=\sum_\lambda\int_0^\infty\frac{\d\omega}{\sqrt{4\pi\omega}}\left[E_I^\lambda(\Omega)a_\lambda(\omega,\Omega)\e^{-\i\omega u}+E_I^{\lambda *}(\Omega)a_\lambda^\dagger(\omega,\Omega)\e^{\i\omega u}\right].
\label{p-form-boundary-modes}
\end{align}
The oscillator algebra is
\begin{align}
[a_\lambda(\omega,\Omega),a_\sigma^\dagger(\omega',\Omega')]=\delta_{\lambda\sigma}\delta(\omega-\omega')\delta_{S^{2p}}(\Omega,\Omega'),
\label{p-form-oscillator-algebra}
\end{align}
with all other commutators equal to zero. Define the matrix of the little-group action by
\begin{align}
(R_h)_{\lambda\sigma}\equiv\frac{1}{p!}E^{*\l I}(\de_hE^\sigma)_I,\qquad R_h^\dagger=-R_h.
\label{p-form-mode-representation}
\end{align}
Substituting \eqref{p-form-boundary-modes} into \eqref{p-form-invariant-generator} and using $\int\d u\,\e^{\i(\omega-\omega')u}=2\pi\delta(\omega-\omega')$, the terms with two annihilation or two creation operators are proportional to $\delta(\omega+\omega')$ and vanish away from the zero mode. The two mixed terms give
\begin{align}
\co_{h}^{(p)}=\i\int_0^\infty\d\omega\int\d\Omega\,:a_\lambda^\dagger(\omega,\Omega)(R_h)_{\lambda\sigma}a_\sigma(\omega,\Omega):.
\label{p-form-mode-generator}
\end{align}
The matrix $\i R_h$ is Hermitian, so this expression is formally self-adjoint.

The number-difference statement follows directly from \eqref{p-form-mode-generator}. For each fixed direction $\Omega$, choose the polarization basis $E_I^\lambda$ already used in \eqref{p-form-polarization-normalization} to diagonalize the Hermitian matrix $\i R_h$. Since $\de_h$ is real and skew-adjoint, its nonzero eigenvalues occur in pairs
\begin{align}
\i \de_hE^\lambda=h_\lambda E^\lambda,\qquad \i \de_hE^{\bar\lambda}=h_{\bar\lambda} E^{\bar\lambda},\qquad h_{\bar\lambda}=-h_\lambda,\qquad h_\lambda\in\mathbb R.
\end{align}
Here the bar denotes the complex-conjugate polarization $E^{\bar\lambda}=E^{*\lambda}$. The flux \eqref{p-form-mode-generator} therefore becomes
\begin{align}
\co_{h}^{(p)}=\int_0^\infty\d\omega\int\d\Omega\sum_{h_\lambda>0}h_\lambda(\Omega)\left[:a_\lambda^\dagger(\omega,\Omega)a_\lambda(\omega,\Omega)-a_{\bar\lambda}^\dagger(\omega,\Omega)a_{\bar\lambda}(\omega,\Omega):\right].
\end{align}
Thus, at every propagation direction $\Omega$, each positive polarization is paired with its opposite polarization, and the flux measures their occupation-number difference. Modes with $h_\lambda=0$ do not contribute.

At the future null infinity, $\Omega=\widehat{\bm k}$ labels the momentum direction. In terms of the bulk oscillators, we obtain
\begin{align}
\co_{h}^{(p)}=\int\d^{2p+1}k\sum_{h_\lambda>0}h_\lambda(\widehat{\bm k})\left[:b_\lambda^\dagger(\bm k)b_\lambda(\bm k)-b_{\bar\lambda}^\dagger(\bm k)b_{\bar\lambda}(\bm k):\right].
\end{align}
The mode argument applies to the nonzero-frequency sector, assumes a positive-frequency splitting and a normal-ordering vacuum, and uses the endpoint condition in \eqref{p-form-hamiltonian-check}. If $h_{AB}(\Omega)$ is angle dependent, a diagonalizing polarization frame need exist only locally, while the invariant expression \eqref{p-form-mode-generator} remains the global definition. A generic $h_{AB}$ defines a transverse $\mathfrak{so}(2p)$ polarization rotation and a weighted difference of conjugate-polarization occupations, not a basis-independent single helicity number.

Finally, $d=2p+2$ makes the bulk Hodge duality of the field strength $H_{p+1}$ have the same degree, and the spherical Hodge duality maps the radiative $p$-form to another $p$-form. Nevertheless, degree matching does not always give a Maxwell-type compact duality. We consider the square of the bulk and sphere Hodge dualities
\begin{align}
(*_d)^2=(-1)^{(p+1)^2+1},\qquad
(*_{2p})^2=(-1)^{p^2}.
\end{align}
For odd $p$, these operators square to $-1$ and define the familiar continuous duality complex structure. For even $p$, they square to $+1$ and instead split the field into real self-dual and anti-self-dual sectors. The general helicity parameters $h_{AB}$ generate $\mathfrak{so}(2p)$ on the transverse indices and should not be identified with the single Hodge-star operation.


\section{Conclusions}
We introduced a metric-compatible covariant variation $\Delta_X=\mathcal L_X+\rho(S_X)=\nabla_X+\rho(A_X)$ on ordinary spacetime tensors. It is the metric Lie derivative determined by the metric lift, and therefore preserves the metric, contractions, the volume form, and Hodge duality. The commutator closes only up to the anomaly $\ca_{X,Y}=-[S_X,S_Y]$, whose Jacobi identity is the corresponding Bianchi-type identity.

The anomaly for a one-form evaluated on a three-dimensional null hypersurface is a helicity rotation, which concerns the electromagnetic duality and the second Chern character. Except the Maxwell field at null infinity in the Motivation, we also consider one-form near a finite horizon and arbitrary tensor fields (especially $p$-form) at null infinity in general dimensions. In both cases, the helicity flux is derived, although only in four dimensions, there is a perfect interpretation regarding the duality and topological term.

We list some further extensions in the following:
\begin{itemize}
\item Other corners in the flat spacetime. We consider the timelike infinity for massive fields with spin 0, 1, and 2 in \cite{Liu:2025oom}. Instead of helicity, the anomaly is generated by a spin charge. One can also explore the covariant variation and its anomaly at the spacelike infinity.
\item Generalization to the anti-de Sitter or de Sitter spacetime. For example, we could consider a Proca field in AdS$_3$. The matter field now has two branches $\ca_a$ and $\cb_a$ near the boundary. Our exploration found that the charge formed by both branches acts on the fields as the covariant variation, and the anomaly leads to an operator of the form below
\begin{align}
m\int\d^2x\,\ep^{ab}\ca_a\cb_b.
\end{align}
This problem deserves more investigation.
\item Topological interpretation beyond one-form and four dimensions. For an Abelian $p$-form potential $B$ in $d=2p+2$ with $H=\d B$, graded commutativity implies $H\wedge H=0$ for even $p$, whereas for odd $p$ it may be nonzero and is locally exact, $H\wedge H=\d(B\wedge H)$. In the latter case, $B\wedge H$ is a Chern-Simons-type transgression form and, after specifying the hypersurface, radiative falloffs, gauge conditions, and endpoint terms, its pullback can generate the Hodge-duality charge. This agrees with the fact that a continuous $\SO(2)$ duality with a Chern-Simons generator exists for $d=4k$, while a single non-chiral field in $d=4k+2$ has no analogous continuous canonical generator \cite{Deser:1997mz,Bandos:2020hgy}. For $p=1$, the reduction to future null infinity reproduces the electromagnetic helicity flux. In general-dimensional Maxwell theory, selected helicity generators can instead be obtained from $F\wedge F\wedge G=\d(A\wedge F\wedge G)$, where $G$ is a closed $(d-4)$-form \cite{Liu:2024rvz}. For $p>1$, however, $B\wedge H$ selects only the Hodge-duality operation and does not reproduce the full $\mathfrak{so}(2p)$-valued family $\co_h^{(p)}$. Realizing a general $h_{AB}$ would require additional background tensors or extra fields and would define a structure beyond $H\wedge H$ alone. Thus, the relation between a topological term and helicity is boundary- and theory-dependent \cite{Long:2025fbb}.

\item Stress-tensor generation of boundary covariant variation. The examples in \cite{Liu:2023qtr,Liu:2024nkc,Liu:2025oom} motivate the following conjecture. Let $X$ be a bulk vector field, and let $\widehat X$ denote the vector field induced on a hypersurface $\ch$. Consider a field theory for which the field $\vp$ reducing to $\ch$ gives the free data $\widehat\vp$. We conjecture that the matter charge or flux
\begin{align}
Q_{\widehat X}=\int_{\ch}(\d^{d-1}x)_\mu\,T^{\mu\nu}X_\nu
\end{align}
generates some transformation of the corresponding field 
\begin{align}
[\i Q_{\widehat X},\widehat\vp]=\check\Delta_{\widehat X}\widehat\vp.
\end{align}
If this $Q_{\wh X}$ is quadratic at the $\wh \vp$, the aforementioned $\check\Delta_{\widehat X}\widehat\vp$ may be the covariant variation or its density-related extension. This relation has been checked for the superrotation at null or timelike infinity in the flat spacetime, but its general validity remains open. A natural future direction is to test it systematically for different spacetimes, hypersurfaces, bulk vector fields, boundary conditions, and matter theories, and to determine whether boundary, corner, or improvement terms are required in $Q_{\widehat X}$. This would clarify how a charge or flux constructed from the bulk stress tensor descends to a generator of covariant variation on the boundary phase space. Also, exploring this may lead us to more applications of the covariant, especially the physical meaning of the anomaly.

\item First-order gravity and mixed fields. The Kosmann derivative has already been used in first-order gravity to construct Lorentz-covariant Noether charges and black-hole entropy, including systems with Majorana and Rarita--Schwinger fields \cite{Jacobson:2015uqa,Prabhu:2015vua,Aneesh:2020fcr,Elgood:2020svt,Oliveri:2020xls}. The total covariant variation defined in section \ref{sec total} may be useful when a field carries both spacetime and Lorentz indices, because it preserves the vielbein $\pl_Xe^a{}_\mu=0$ for every $X$.  We have not found the exact mixed-index packaging \eqref{total-covariant-variation} like our total covariant variation, although its ingredients overlap with metric or reductive Lie derivative. We hope to investigate its potential applications.
\end{itemize}

\acknowledgments The work of J.L. is supported by NSFC Grant No. 12575074.

\appendix
\section{Covariant variation of differential forms with higher rank}\label{higher forms}
In our convention, the volume form takes the expression
\begin{align}
\bm\ep&=\frac{1}{d!}\epsilon_{\m_1\cdots\m_d}\d x^{\mu_1}\wedge\cdots \wedge \d x^{\mu_d}
\end{align}
with $\epsilon_{\m_1\cdots\m_d}=\sqrt{|g|}\widetilde\epsilon_{\m_1\cdots\m_d}$ and $\widetilde\epsilon_{01\cdots d-1}=1$. Tensors (forms) with rank 0, 1, 2 are denoted by usual characters, while their Hodge duals are denoted by bold characters. For example, 
\begin{subequations}
\begin{align}
*L&=L\bm\ep=L(\d^dx)=\bm L, \\ *\th&=\th\cdot\bm\ep=\th^\m(\d^{d-1}x)_\m=\bm\th,\\
*q&=q^\mn(\d^{d-2}x)_\mn=\bm q,
\end{align}
\end{subequations}
are the differential forms  associated with the Lagrangian, presymplectic potential, and surface charge, respectively. Our conventions are
\begin{subequations}
\begin{align}
(*\omega)_{\mu_1 \cdots \mu_{d-p}}&=\frac{1}{p!} \epsilon^{\nu_1 \cdots \nu_p}{}_{\mu_1 \cdots \mu_{d-p}} \omega_{\nu_1 \cdots \nu_p},\\
\big(\d^{d-p} x\big)_{\mu_1 \cdots \mu_p} &= \frac{1}{p!  (d-p)!}  \epsilon_{\mu_1 \cdots \mu_d} \d x^{\mu_{p+1}} \wedge \cdots \wedge \d x^{\mu_d}
\end{align}
\end{subequations}
such that
\begin{align}
*\omega=\bm\omega=\omega^{\mu_1 \cdots \mu_p}(\d^{d-p}x)_{\mu_1 \cdots \mu_p}.
\end{align}

The Lie derivative of $\bm{\omega}$ reads
\begin{align}
\mathcal{L}_X \bm{\omega}=\mathcal{L}_X \omega^{\mu_1 \cdots \mu_p} (\d^{d-p}x)_{\mu_1 \cdots \mu_p} + \bm{\omega}\nabla \cdot X,
\end{align}
where we have used
\begin{align}
\mathcal{L}_X (\d^{d-p}x)_{\mu_1 \cdots \mu_p}=\nabla \cdot X(\d^{d-p}x)_{\mu_1 \cdots \mu_p}.
\end{align}
Here, \((\mathrm d^{d-p}x)_{\mu_1\cdots\mu_p}\in\Lambda^{d-p}(M,\Lambda^pT^*M)\), i.e., it is $\Lambda^pT^*M$-valued $(d-p)$-form, and the Lie derivative acts on both form and label indices. Now we consider the covariant variation
\begin{align}
\Delta_X\bm\omega
&= \Big[\mathcal{L}_X \omega^{\mu_1 \cdots \mu_p} + \sum_{k=1}^p (S_X)^{\mu_k}_\n \omega^{\mu_1 \cdots \n\cdots \mu_p}\Big] (\d^{d-p}x)_{\mu_1 \cdots \mu_p},
\end{align}
where we have used
\begin{align}
\Delta_X(\d^{d-p}x)_{\mu_1 \cdots \mu_p}=0.
\end{align}
Substituting the Lie derivative, we obtain
\begin{align}
\Delta_X \bm{\omega} = \mathcal{L}_X \bm{\omega} - \bm{\omega}\nabla \cdot X +p (S_X)^{\mu_1}_\n \omega^{\nu\mu_2 \cdots \mu_p} (\d^{d-p}x)_{\mu_1 \cdots \mu_p}.
\end{align}

If $p=0$, we get the formula for a top form $\bm L=L\bm\ep$ 
\begin{align}
\Delta_X \bm{L} = \mathcal{L}_X \bm{L} - \bm{L}\nabla \cdot X.
\end{align}
In particular, we find
\begin{align}
\Delta_X\bm\ep=0,
\end{align}
which is equivalent to
\begin{align}
\Delta_X \epsilon_{\m_1\cdots\m_d}=X^\r\nabla_\r \epsilon_{\m_1\cdots\m_d}+\sum_{i=1}^d \nabla_{[\m_i}X_{\r]}\epsilon_{\m_1\cdots}{}^\rho{}_{\cdots\m_d}=\tr(\nabla_{[\m}X_{\n]})\epsilon_{\m_1\cdots\m_d}=0.
\end{align}
The other interesting cases are $p=1$ and 2, for which we obtain
\begin{align}
\Delta_X \bm{\theta} &= \mathcal{L}_X \bm{\theta} - \bm{\theta}\nabla \cdot X +(S_X)^{\mu}_\n \theta^\n (\d^{d-1}x)_\mu,\\
\Delta_X \bm{Q} &= \mathcal{L}_X \bm{Q} - \bm{Q}\nabla \cdot X +2(S_X)^{\mu}_\r Q^{\r\nu} (\d^{d-2}x)_{\mu\nu}.
\end{align}

For a top form $\bm L$, we can compute 
\begin{align}
[\Delta_X,\Delta_Y]\bm{L}&=\cl_X(\mathcal{L}_Y \bm{L} - \bm{L} \nabla\cdot Y)-(\mathcal{L}_Y \bm{L} - \bm{L} \nabla\cdot Y)\nabla\cdot X-(X\leftrightarrow Y)\nn\\
&=\mathcal{L}_{[X,Y]} \bm{L}+\bm L[\cl_Y(\nabla\cdot X)-\cl_X(\nabla\cdot Y)]\nn\\
&=\mathcal{L}_{[X,Y]} \bm{L}-\bm L\nabla \cdot [X,Y],
\end{align}
where we have used the identity
\begin{align}
\mathcal{L}_X(\nabla \cdot Y) - \mathcal{L}_Y(\nabla \cdot X) = \nabla \cdot [X,Y].
\end{align}
Thus, the anomaly is
\begin{align}
\ca_{X, Y}\bm L&=[\Delta_X,\Delta_Y]\bm{L}-\Delta_{[X,Y]} \bm{L}=0,
\end{align}
which is consistent with the anomaly formula \eq{general anomaly}, since
\begin{align}
\mathcal{A}_{X,Y} \ep_{\mu_1 \cdots \mu_d} &= \frac{1}{4} \sum_{j=1}^d \left[ \mathcal{L}_X g_{\mu_j \rho} \mathcal{L}_Y g^{\rho\sigma} - (X \leftrightarrow Y) \right] \ep_{\mu_1 \cdots \sigma \cdots \mu_d}
=0,
\end{align}
and also, $\Delta_X\bm\ep=0$ of course will lead to $\mathcal{A}_{X,Y}\bm\ep=0$. In general, the anomaly vanishes for a scalar or a top form.

\section{Action on connection and Riemann tensor}
\subsection{Case of Lie derivative}\label{Lie conn}
Under a diffeomorphism induced by $X$, we have 
\begin{align}
  x'^\m=x^\m+\epsilon X^\m \qrq A'_\m(x')=\frac{\p x^\n}{\p x'^\m}A_\n(x)=A_\m-\epsilon\p_\m X^\n A_\n+O(\epsilon^2),
\end{align}
and then the Lie derivative of a one-form is defined as
\begin{align}
  \cl_X A_\m&=\lim_{\epsilon\to0}\frac{1}{\epsilon}[A_\m(x)-A'_\m(x)]\nn\\
  &=\lim_{\epsilon\to0}\frac{1}{\epsilon}[A_\m(x)-A'_\m(x')+\epsilon X^\n\p_\n A_\m(x)+O(\epsilon^2)]\nn\\
  &=X^\n\p_\n A_\m+A_\n\p_\m X^\n.\label{clxia}
\end{align}
For a vector, we have
\begin{align}
  \cl_X V^\m&=\lim_{\epsilon\to0}\frac{1}{\epsilon}[V^\m(x)-V'^\m(x)]=X^\n\p_\n V^\m-V^\n\p_\n X^\m.\label{clxiv}
\end{align}

Under $x'=x+\epsilon X$, the partial derivative of a vector transforms as 
\begin{align}
  \p'_\m V'^\n(x)&=\frac{\p x^\r}{\p x'^\m}\p_\r\Big[\frac{\p x'^\n}{\p x^\s}V^\s\Big]-\ep X^\r\p_\r\p_\m V^\n+O(\ep^2)\nn\\
  &=\p_\m V^\n-\ep\p_\m(X^\r\p_\r V^\n-V^\r\p_\r X^\n)+O(\ep^2),
\end{align}
and with the previous definition, we find
\begin{align}
\cl_{X}(\p_\m V^\n)=\lim_{\epsilon\to0}\frac{1}{\epsilon}[\p_\m V^\n-\p_\m' V'^\n]=\p_\m[X,V]^\n=\p_\m(\cl_{X}V^\n).
\end{align}
The same can be generalized to all the tensors. Then we conclude $[\cl_X,\p_\m]=0$. It is correct in the sense that $\partial_\mu V^\nu$ is a jet coordinate. We prolong an active field-space variation $\cl_XV^\n$ at fixed coordinates to the jet bundle by
\begin{align}
\delta_X^{\mathrm{pr}}V^\nu=\mathcal{L}_XV^\nu, \qquad \delta_X^{\mathrm{pr}}(\partial_{\mu_1}\cdots\partial_{\mu_k}V^\nu)=\partial_{\mu_1}\cdots\partial_{\mu_k}(\delta_X^{\mathrm{pr}}V^\nu). \label{jet-prolongation}
\end{align}
Therefore, by definition we have
\begin{align}
[\delta_X^{\mathrm{pr}},\partial_\mu]=0.\label{jet-commutation}
\end{align}
On the other hand, if we see $\partial_\mu=e_\m$ itself as a vector, then $e_\m[f]=\p_\m f$ is a function like the usual $X[f]$. Then
\begin{align}
[\mathcal{L}_X,\partial_\mu]f =-(\partial_\mu X^\n)\partial_\n f. \label{geometric-partial-commutator}
\end{align} 
Equivalently, this is
\begin{align}
[\mathcal L_X,\partial_\mu]f=(\mathcal L_X\partial_\mu)[f]=[X,\partial_\mu][f].
\end{align}
Here the coordinate basis itself is also Lie dragged, while the jet prolongation keeps the derivative label \(\mu\) fixed by definition. In summary, the first way is to see $\m$ in $\p_\m V^\n$ as a tensor index like other indices,\footnote{It is common in the physics to use the components to represent the tensor itself. For example, we use $\cl_XV^\m\equiv(\cl_XV)^\m$ to represent $\cl_XV$, and use $\nabla_\m V^\n=(\nabla V)_\m{}^\n$ to represent $\nabla V$.} although $\p_\m V^\n$ is not a tensor; while in the second way, $\m$ is just a label to distinguish the different basis vectors.
 
To complete the picture, from the transformation law of connection
\begin{align}
\Gamma_{\rs}'^\m(x')&=\frac{\p x'^\m}{\p x^\n}\frac{\p x^\k}{\p x'^\r}\frac{\p x^\l}{\p x'^\s}\Gamma_{\k\l}^\n-\frac{\p x^\k}{\p x'^\r}\frac{\p x^\l}{\p x'^\s}\frac{\p^2 x'^\m}{\p x^\k\p x^\l}\nn\\
&=\Gamma_{\rs}^\m+\ep(\Gamma_{\rs}^\n\p_\n X^\m-\Gamma_{\n\s}^\m\p_\r X^\n-\Gamma_{\r\n}^\m\p_\s X^\n-\p_\r\p_\s X^\m)+O(\ep^2),\label{ksfbhgl}
\end{align}
we can compute
\begin{align}
\cl_X\Gamma_{\rs}^\m&=\lim_{\ep\to0}\frac{1}{\ep}[\Gamma_{\rs}^\m(x)-\Gamma_{\rs}'^\m(x)]\nn\\
&=X^\n\p_\n\Gamma_{\rs}^\m-\Gamma_{\rs}^\n\p_\n X^\m+\Gamma_{\n\s}^\m\p_\r X^\n+\Gamma_{\r\n}^\m\p_\s X^\n+\p_\r\p_\s X^\m\label{lxig1}\\
&=\nabla_{(\r}\nabla_{\s)}X^\m-R^\m{}_{(\r\s)\n}X^\n,\label{lxig}
\end{align}
which agrees with the result from taking Lie derivative in the general variation of connection, i.e.,
\begin{align}
\delta\Gamma^\mu_{\rho\sigma}=\frac{1}{2}g^{\mu\nu}(\nabla_\rho \delta g_{\sigma\nu}+\nabla_{\sigma}\delta g_{\rho \nu}-\nabla_\nu \delta g_{\rho\sigma}).\label{deGamma}
\end{align}

From the fact $[\cl_X,\p_\m]=0$, we find
\begin{align}
\cl_X(\nabla_\m A_\n)&=\cl_X(\p_\m A_\n-\Gamma^\r_\mn A_\r)\nn\\
&=\p_\m(\cl_X A_\n)-\cl_X\Gamma^\r_\mn A_\r-\Gamma^\r_\mn\cl_X A_\r\nn\\
&=\nabla_\m(\cl_X A_\n)-\cl_X\Gamma^\r_\mn A_\r,\label{LxinablaA}
\end{align}
and more generally we have
\begin{align}
[\cl_X,\nabla_\r]T^\m{}_\n=\cl_X\Gamma_{\r\s}^\m T^\s{}_\n-\cl_X\Gamma_{\r\n}^\s T^\m{}_\s. \label{Lxinabla}
\end{align}
As said, this can be seen as the definition for Lie derivative of connection. This construction and the corresponding curvature identities go back to the classical treatments of Lie derivatives of connections and infinitesimal affine transformations \cite{MR0088769,AIF_1964__14_2_227_0,MR0152974,KolarMichorSlovak1993,doi:10.1142/S0219887825500859}.

The Lie derivative of Riemann tensor of course reads
\begin{align}
\cl_XR^\rho{}_{\sigma\mu\nu}
&=X^\lambda\nabla_\lambda R^\rho{}_{\sigma\mu\nu}
-R^\lambda{}_{\sigma\mu\nu}\nabla_\lambda X^\rho
+R^\rho{}_{\lambda\mu\nu}\nabla_\sigma X^\lambda\nn\\
&\quad
+R^\rho{}_{\sigma\lambda\nu}\nabla_\mu X^\lambda
+R^\rho{}_{\sigma\mu\lambda}\nabla_\nu X^\lambda.\label{Lie-Riemann}
\end{align}
For a Killing vector $X$, the identity $\nabla_{\mu} \nabla_{\nu} X_{\rho} = R_{\rho\nu\mu\sigma} X^{\sigma}$ ensures the vanishing of $\cl_X R^\rho{}_{\sigma\mu\nu}$. If we introduce $h_{\mu\nu} = \mathcal{L}_{X} g_{\mu\nu} = 2 \nabla_{(\mu} X_{\nu)}$, it can also be written as
\begin{align}
\mathcal{L}_{X} R^{\rho}{}_{\sigma\mu\nu} = \frac{1}{2} g^{\rho\lambda} \nabla_{\mu} ( \nabla_{\sigma} h_{\nu\lambda} + \nabla_{\nu} h_{\sigma\lambda} - \nabla_{\lambda} h_{\sigma\nu}) - (\mu \leftrightarrow \nu),\label{clxiR}
\end{align}
which implies 
\begin{align}
\mathcal{L}_{X} g_{\mu\nu}=0 \qrq \mathcal{L}_{X} R^{\rho}{}_{\sigma\mu\nu}=0.
\end{align}
As a check, on the flat background, \eq{clxiR} is equivalent to the famous linearized Riemann tensor
\begin{align}
\cl_X R_{\mu\nu\rho\sigma} =\frac{1}{2}(\partial_\m\partial_\s h_{\nu\r}-\partial_\n\partial_\s h_{\mu\r}-\partial_\m\partial_\r h_{\nu\s}+\partial_\n\partial_\r h_{\mu\s}).
\end{align}

\subsection{Covariant variation of connection}\label{CVch}
An affine connection is not a tensorial associated field, so the slotwise definition of $\Delta_X$ does not apply to it. The following formulas are additional conventions and must be tested separately for the covariance, metric compatibility, and curvature. The commutator between the covariant variation and covariant derivative is
\begin{align}
[\Delta_X,\nabla_\mu]V^\nu&=\Delta_X(\p_\m V^\n+\Gamma^\n_{\m\r}V^\r)-(\p_\m\Delta_X V^\n+\Gamma^\n_{\m\r}\Delta_X V^\r)\nn\\
&=[\Delta_X,\p_\m]V^\n+(\Delta_X\Gamma^\n_{\m\r})V^\r.
\end{align}
It cannot uniquely determine a covariant variation of the connection. In the fixed-label convention, $[\cl_X,\p_\m]$ vanishes and $\cl_X\Gamma^\n_{\m\r}$ is determined by the commutator. For $\Delta_X$, however, $[\Delta_X,\p_\m]V^\n\ne0$, since the explicit computation shows
\begin{align}
[\Delta_X,\nabla_\mu]V^\nu&=X^\rho[\nabla_\rho,\nabla_\mu] V^\nu-\nabla_{(\mu}X_{\rho)}\nabla^\rho V^\nu -V_\rho\nabla_\mu\nabla^{[\nu}X^{\rho]},\label{eq429}
\end{align}
where the second term depends on the derivative of $V$.

Only the sum in the commutator is fixed. After a tensorial convention for the connection variation has been chosen, the action of covariant variation on $\p_\m V^\n$ is determined
\begin{align}
\Delta_X(\p_\m V^\n)&=[\Delta_X,\nabla_\mu]V^\nu-(\Delta_X\Gamma^\n_{\m\r})V^\r+\p_\m(\Delta_X V^\n).
\end{align}
Below, we discuss several choices for the covariant variation of the connection:
\begin{enumerate}
\item For the Levi-Civita connection, since $\Gamma^\n_{\m\r}$ is symmetric about two lower indices, and the correction term for covariant variation is induced by $S_X$, we may choose
\begin{align}
\Delta_X\Gamma^\n_{\m\r}&= \cl_X\Gamma^\n_{\m\r}-\frac{1}{2}\nabla^\n(\cl_X g_{\m\r}).
\end{align}
\item We may also take
\begin{align}
\Delta_X\Gamma^\n_{\m\r}&= \cl_X\Gamma^\n_{\m\r}-\nabla_\m (S_X)^\n_\r,
\end{align}
which will be discussed in appendix \ref{gra as gauge conn}. If restricted to the Levi-Civita connection, one may symmetrize the second term by hand
\begin{align}
\Delta_X\Gamma^\n_{\m\r}&= \cl_X\Gamma^\n_{\m\r}-\nabla_{(\m} (S_X)^\n_{\r)}.
\end{align}
\item Since the Lie derivative of connection has already been a tensor, a natural choice may be to take
\begin{align}
\Delta_X\Gamma^\n_{\m\r}&= \cl_X\Gamma^\n_{\m\r}= \nabla_{(\m}\nabla_{\r)}X^\n-R^\n{}_{(\m\r)\s}X^\s.
\end{align}
\item The simplest choice is to take
\begin{align}
\De\Gamma^\n_{\m\r}=0.
\end{align}
\end{enumerate}

{
\paragraph{Consistency checks.}
Given a choice of the covariant variation of connection $\Delta\Gamma$, $\De g=0$, and
\begin{align}
\Delta_X R^\rho{}_{\sigma\mu\nu} &= \mathcal{L}_X R^\rho{}_{\sigma\mu\nu} +(S_X)^\r_\l R^\lambda{}_{\sigma\mu\nu} - (S_X)^\l_\s R^\rho_{\l\mu\nu} - (S_X)^\l_\m R^\rho{}_{\sigma\l\n}- (S_X)^\l_\n R^\rho{}_{\sigma\mu\l},\label{DeXR}
\end{align}
it is natural to do two consistency checks. The first one is to directly take covariant variation of
\begin{align}
\Gamma^\r_{\mn}=\frac12g^{\rs}(\partial_\m g_{\n\s}+\partial_\n g_{\m\s}-\partial_\s g_{\mn}),\label{eqB34}
\end{align}
and see whether it leads to the chosen $\Delta\Gamma$. Since the covariant variation kills metric, we find
\begin{align}
\Delta_X \Gamma^\rho_{\mu\nu} = \frac{1}{2} g^{\rho\sigma} [ \Delta_X(\partial_\mu g_{\nu\sigma}) + \Delta_X(\partial_\nu g_{\mu\sigma}) - \Delta_X(\partial_\sigma g_{\mu\nu})].\label{eqb30}
\end{align}
Here $\Delta_X(\partial_\mu g_{\nu\sigma})$ can be computed from
\begin{align}
0&=\Delta_X(\nabla_\mu g_{\nu\sigma})=\Delta_X(\partial_\mu g_{\nu\sigma}-\Gamma^\r_\mn g_\rs-\Gamma^\r_{\m\s}g_{\n\r}),
\end{align}
and the result is
\begin{align}
\Delta_X(\partial_\mu g_{\nu\sigma}) = \Delta_X\Gamma^\r_\mn g_\rs+\Delta_X\Gamma^\r_{\m\s}g_{\n\r}.\label{DelXpg}
\end{align}
Substituting \eq{DelXpg} back into \eq{eqb30} gives no constraint. 

The second check is to directly take covariant variation of
\begin{align}
[\nabla_\mu, \nabla_\nu] V^\rho = R^\rho{}_{\sigma\mu\nu} V^\sigma,\label{Riem def}
\end{align}
and compare the result with \eq{DeXR}. From the Jacobi identity
\begin{align}
[\Delta_X,[\nabla_\mu,\nabla_\nu]]=[[\Delta_X,\nabla_\mu],\nabla_\nu]+[\nabla_\mu,[\Delta_X,\nabla_\nu]],
\end{align}
the covariant variation of the curvature actually only relies on the sum $[\Delta_X,\nabla_\mu]$. Hence this check cannot select a connection variation.

\paragraph{A family for covariant variation of connection.}
In general, we can define
\begin{align}
\Delta_X\Gamma^\rho_{\mu\nu}=(P_X)^\rho_{\mu\nu}+(C_X)^\rho_{\mu\nu},
\label{connection-C-family}
\end{align}
where $(P_X)^\rho_{\mu\nu}\equiv(\mathcal L_X\Gamma)^\rho_{\mu\nu}$, and $C_X$ is to be determined. For the Levi-Civita connection,
\begin{align}
(P_X)^\rho_{\mu\nu}
=\nabla_\mu(S_X)^\rho{}_\nu+\nabla_\nu(S_X)^\rho{}_\mu-\nabla^\rho(S_X)_{\mu\nu}.
\label{P-in-S}
\end{align}
On the other hand, the tensor operator determines
\begin{align}
[\Delta_X,\nabla_\mu]V^\rho
=-(S_X)_\mu{}^\n\nabla_\n V^\rho
+(Q_X)^\rho{}_{\mu\n}V^\n,
\label{fixed-Delta-nabla}
\end{align}
where $Q_X=\cl_X\nabla-\nabla S_X$ with the following component
\begin{align}
(Q_X)^\rho_{\mu\nu}
=\nabla_\nu(S_X)^\rho{}_\mu-\nabla^\rho(S_X)_{\mu\nu}
=X^\s R^\rho{}_{\nu\s\mu}
-\nabla_\mu(A_X)^\rho{}_\nu.
\end{align}
Equations \eqref{connection-C-family} and \eqref{fixed-Delta-nabla} are compatible provided that the rule for the partial derivative changes with $C_X$ as
\begin{align}
[\Delta_X,\partial_\mu]V^\rho
=-(S_X)_\mu{}^\s\nabla_\s V^\rho
-\left[\nabla_\mu(S_X)^\rho{}_\sigma+(C_X)^\rho{}_{\mu\sigma}\right]V^\sigma.
\label{C-dependent-partial-rule}
\end{align}


More explicitly, the choices above lie in a family that is linear in the first derivative of $S_X$,
\begin{align}
\Delta_X^{(c_1,c_2,c_3)}\Gamma^\rho_{\mu\nu}=(P_X)^\rho_{\mu\nu}+c_1\nabla_\mu(S_X)^\rho{}_\nu+c_2\nabla_\nu(S_X)^\rho{}_\mu+c_3\nabla^\rho(S_X)_{\mu\nu}.
\label{minimal-connection-family}
\end{align}
Choices 1, 2, its symmetrized form, 3, and 4 correspond respectively to
\begin{align}
(c_1,c_2,c_3)
=(0,0,-1),\quad(-1,0,0),\quad
(-\frac12,-\frac12,0),\quad (0,0,0), \qaq (-1,-1,1).
\end{align}
Allowing trace gradients or curvature terms enlarges the family further. Within this ansatz, torsion-free condition requires $c_1=c_2$, but there is no other constraint. One must specify the intended object from some additional input. For instance, keeping the similarity as the Lie derivative selects choice 3, and identifying the coefficient variation with the operator variation $\Delta_X\nabla$ selects the unsymmetrized choice 2.  
}

\subsection{Kosmann derivative as a covariant Lie derivative}\label{Kosm as cov Lie}
\paragraph{Covariant Lie derivative in a gauge theory.}
Consider a gauge theory with group $G$, connection $A$, curvature $F=\d A+A\wedge A$, and covariant derivative $D=\d+\rho_G(A)$. The Lie derivative of the gauge potential is easy to find
\begin{align}
\mathcal{L}_X A_\mu &= X^\nu \partial_\nu A_\mu + A_\nu \partial_\mu X^\nu=X^\nu F_{\nu\mu} + D_\mu (X^\nu A_\nu).\label{LxiAmu}
\end{align}

Under a local gauge transformation $U(x)\in G$, the connection and a physical field $\psi$ in the fundamental representation transform as
\begin{align}
A'_\mu = U A_\mu U^{-1} + U \partial_\mu U^{-1} \qaq \psi'=U\psi.
\end{align}
The field strength belongs to the adjoint representation $F'_{\mu\nu} = U F_{\mu\nu} U^{-1}$, and the gauge covariant derivative is designed to satisfy
\begin{align}
D'_\mu = U D_\mu U^{-1} \qrq (D_\mu \psi)' = U (D_\mu \psi).
\end{align}

In \eq{LxiAmu}, the first part is gauge covariant under the adjoint representation, while the second part is not. We modify it to be a covariant Lie derivative
\begin{align}
\wt\cl_X A_\mu \equiv X^\nu F_{\nu\mu} + D_\mu (X^\nu A_\nu - K_X) ,
\end{align}
where $K_X$ generates an infinitesimal gauge transformation, and is chosen such that
\begin{align}
X^\nu A'_\nu - K'_X = U (X^\nu A_\nu - K_X) U^{-1}.
\end{align}
Solving this, we obtain
\begin{align}
K'_X = U K_X U^{-1} - \mathcal{L}_X U U^{-1}.\label{adj trans}
\end{align}

In a more concise way, the covariant Lie derivative is \cite{Jackiw:1979ub,Obukhov:2006ge,Prabhu:2015vua,Elgood:2020svt,Giotopoulos:2025weu}.
\begin{align}
\wt{\cl}_X = \mathcal{L}_X + \delta_{K_X},
\end{align}
which acts on various fields as
\begin{subequations}
\begin{align}
\wt{\cl}_X \psi&=\cl_X\psi+K_X \psi,\\
\wt{\cl}_X A_\m&=\cl_X A_\m-D_\m K_X,\\
\wt{\cl}_X F_{\mu\nu} &= \mathcal{L}_X F_{\mu\nu} + [K_X, F_{\mu\nu}],\\
\wt{\cl}_X (D_\m\psi)&=\cl_X(D_\m\psi)+K_X D_\m\psi.
\end{align}
\end{subequations}

For completeness, the commutator with a directional covariant derivative contains both the transported direction and the connection variation
\begin{align}
[\wt{\mathcal{L}}_X,D_Y]\psi =D_{[X,Y]}\psi +\rho_G\left((\wt{\mathcal{L}}_XA)(Y)\right)\psi. \label{gauge-directional-commutator}
\end{align}
In a coordinate basis, this reads
\begin{align}
[\wt{\mathcal{L}}_X,D_\mu]\psi =-(\partial_\mu X^\nu)D_\nu\psi+\rho_G(\wt{\mathcal{L}}_XA_\mu)\psi. \label{gauge-coordinate-commutator}
\end{align}
Dropping the first term amounts to using the fixed-coordinate prolongation \eqref{jet-commutation} instead of the geometric commutator.

\paragraph{Lorentz group as gauge group.}
The Kosmann derivative is actually a covariant Lie derivative under the local Lorentz group. Under a Lorentz transformation $\L$, we have
\begin{align}
e'^a_\mu = \Lambda^a{}_b e^b_\mu \quad \text{and} \quad e_a'^\mu =(\L^{-1})^b{}_ae^\mu_b =\L_a{}^b e^\mu_b.
\end{align}
Recalling $(\lambda_X)^{ab} = e^{\mu[a} \mathcal{L}_X e^{b]}_\mu$, then
\begin{align}
\lambda'_X=\Lambda \lambda_X \Lambda^{-1} + \Lambda \mathcal{L}_X \Lambda^{-1},
\end{align}
i.e., $\l_X$ satisfies the above property \eqref{adj trans}.

Given that $\omega^{ab}$ is the gauge potential for the Lorentz group, we can define the Kosmann derivative of the spin connection
\begin{align}
\mathcal{K}_X \omega^{ab} &= \cL_X\omega^{ab} - D(\l_X)^{ab} = i_X R^{ab} + D(i_X\omega -\l_X)^{ab},\label{Kdiffomega2}
\end{align}
where the ordinary Cartan identity applied to the connection one-form gives
\begin{align}
\mathcal L_X\omega^{ab}&=i_X\d\omega^{ab}+\d(i_X\omega^{ab})\nn\\
&=i_XR^{ab}+Di_X\omega^{ab}.
\end{align}
On the other hand, the Kosmann derivative acts on a matter field $\psi$ as
\begin{align}
\mathcal{K}_X\psi &=X^\mu D_\mu\psi+\rho_{\rm L}(A_X)\psi.
\end{align}

\subsection{Covariant variation as a covariant Lie derivative?}\label{gra as gauge conn}
If we introduce the generators of the general linear group
\begin{align}
  (G_\r{}^\s)^\n{}_\m=\i\de^\s_\m \de^\n_\r \qaq ( G_\r{}^\s)_\n{}^\m=-\i\de^\s_\n \de^\m_\r, \label{Jsrnm}
\end{align}
then we can rewrite the covariant derivative as
\begin{align}
  \nabla_\mu = \partial_\mu + \Gamma_\mu= \partial_\mu - \i\,\Gamma^\rho_{\mu\sigma}  G_\r{}^\s.
\end{align}
It is easy to find
\begin{align}
  &\Gamma_\mu V^\n=- \i\,\Gamma^\rho_{\mu\sigma}( G_\r{}^\s)^\n{}_\l V^\l= \Gamma^\rho_{\mu\sigma}\delta^\n_{\r} V^\s=\Gamma^\n_{\mu\sigma} V^\sigma,\\ 
  &\Gamma_\mu A_\n = - \i\,\Gamma^\rho_{\mu\sigma}( G_\r{}^\s)_\n{}^\l A_\l= -\Gamma^\rho_{\mu\sigma}\delta^\s_{\n} A_\r = - \Gamma^\r_{\mu\n} A_\r,
\end{align}
and thus
\begin{align}
  &\nabla_\mu V^\n = \partial_\mu V^\n + \Gamma^\n_{\mu\sigma} V^\sigma \qaq \nabla_\mu A_\n  = \partial_\mu A_\n - \Gamma^\r_{\mu\n} A_\r.
\end{align}

The spin connection is the gauge potential of local $\SO(1,d-1)$. Similarly, the spacetime connection coefficients may be assembled into a $\mathfrak{gl}(d,\mathbb R)$-valued one-form
\begin{align}
\Gamma^\n{}_\r=\Gamma^\n_{\m\r}\d x^\m.
\end{align}
We try to define an explicit End$(TM)$-connection variation 
\begin{align}
(Q_X)^\nu{}_{\mu\rho}&\equiv\mathcal L_X\Gamma^\nu{}_{\mu\rho}-\nabla_\mu(S_X)^\nu{}_\rho=(\Delta_X\nabla)_\mu{}^\nu{}_\rho.\label{Q-connection-definition}
\end{align}
However, the metric compensator $S_X$ is an End$(TM)$-valued tensor, not an independent gauge parameter with the inhomogeneous transformation law of a connection compensator. Therefore, the covariant variation cannot be treated as a covariant Lie derivative.

For the Levi-Civita connection, the vielbeins intertwine the Kosmann derivative on Lorentz tensors with the covariant variation on the corresponding spacetime tensors, \(\mathcal K_X\circ e=e\circ\Delta_X\). More explicitly, this is
\begin{align}
\mathcal K_XV^a=\ck_X(e^a_\mu V^\mu)=e_\m^a\Delta_XV^\mu.
\end{align}
This equality does not make \(\Delta_X\) a gauge-covariant Lie derivative. The Kosmann derivative acts on fields naturally associated with the internal Lorentz bundle, whereas ordinary spacetime tensors carry the natural action of diffeomorphisms and are not matter fields of an independent internal gauge theory. The vielbein relates these two actions in the soldered representation, but it does not identify their geometric origins.

\subsection{Covariant variation of Riemann tensor}
Because the difference of two connections is tensorial, the curvature variation generated by $Q_X$ is
\begin{align}
(\widehat{\Delta}_XR)^\rho{}_{\sigma\mu\nu}=2\nabla_{[\mu}(Q_X)^\rho{}_{\nu]\sigma}. \label{Q-curvature}
\end{align}
Equivalently,
\begin{align}
(\widehat{\Delta}_XR)^\rho{}_{\sigma\mu\nu}&=(\mathcal{L}_XR)^\rho{}_{\sigma\mu\nu} +(S_X)^\rho{}_\lambda R^\lambda{}_{\sigma\mu\nu} -(S_X)^\lambda{}_\sigma R^\rho{}_{\lambda\mu\nu}. \label{Q-curvature-internal}
\end{align}
The hat is important: \eqref{Q-curvature-internal} is the variation of the curvature as an $\operatorname{End}(TM)$-valued two-form, so the metric compensator has acted on the two internal indices but not yet on the two form indices. Acting on those form indices as well gives the ordinary-tensor covariant variation,
\begin{align}
(\Delta_XR)^\rho{}_{\sigma\mu\nu}&=(\widehat{\Delta}_XR)^\rho{}_{\sigma\mu\nu} -(S_X)^\lambda{}_\mu R^\rho{}_{\sigma\lambda\nu} -(S_X)^\lambda{}_\nu R^\rho{}_{\sigma\mu\lambda} \label{Delta-R-from-Q}\\
&=X^\lambda\nabla_\lambda R^\rho{}_{\sigma\mu\nu} +(A_X)^\rho{}_\lambda R^\lambda{}_{\sigma\mu\nu} -(A_X)^\lambda{}_\sigma R^\rho{}_{\lambda\mu\nu}\nn\\ 
&\quad -(A_X)^\lambda{}_\mu R^\rho{}_{\sigma\lambda\nu} -(A_X)^\lambda{}_\nu R^\rho{}_{\sigma\mu\lambda}. \label{Delta-R-A}
\end{align}
It is obvious that $\Delta_Xg=0$ does not imply $\Delta_XR=0$.

The condition $\Delta_XR=0$ for every vector field $X$ can be classified pointwise. At a point, $X^\mu$ and $(A_{X})_\mn=\nabla_{[\mu}X_{\nu]}$ can be prescribed independently. Equation \eqref{Delta-R-A} then requires
\begin{align}
\nabla R=0 \qaq \rho(A_X)R=0.
\end{align}
An algebraic curvature tensor invariant under the full orthogonal algebra is proportional to $g_{\mu\rho}g_{\nu\sigma}-g_{\mu\sigma}g_{\nu\rho}$. Therefore,
\begin{align}
\Delta_XR_{\mu\nu\rho\sigma}=0\quad\text{for every }X \quad\Longleftrightarrow\quad R_{\mu\nu\rho\sigma} =K(g_{\mu\rho}g_{\nu\sigma}-g_{\mu\sigma}g_{\nu\rho}), \quad \d K=0, \label{Delta-R-classification}
\end{align}
locally. This is constant sectional curvature. To obtain a maximally symmetric spacetime additionally requires the usual global hypotheses and thus is not a purely local consequence.

Denote the Ricci tensor and Ricci scalar by $\cR_{\mu\nu}$ and $\cR$, respectively. The analogous lower-rank statements are
\begin{align} 
\Delta_X\cR_{\mu\nu}=0\ \text{for every }X &\quad\Longleftrightarrow\quad \cR_{\mu\nu}=\Lambda g_{\mu\nu}, \quad \d\Lambda=0,\\
\Delta_X\cR=0\ \text{for every }X &\quad\Longleftrightarrow\quad \d \cR=0.
\end{align}
For $d\geq3$, the constancy of $\Lambda$ in the first line also follows from the contracted Bianchi identity.

\section{Jacobi identity}\label{sec jacobi}
Given all the commutators in \eq{comm V}, there are $\binom{5}{3}=10$ independent Jacobi identities involving these three operators. We explicitly compute all of them, among which some lead to useful geometric identities\footnote{Note that in the last three lines of \eq{Jacobi ids}, we use ``cyclic'' to mean cyclic permutations of different kinds of objects.}
\begin{subequations}\label{Jacobi ids}
\begin{small}
\begin{align}
&\sum_{\text{cyclic}}[\mathcal{L}_X, [\mathcal{L}_Y, \cl_Z]] V^\mu =0 \qrq \sum_{\text{cyclic}}[X,[Y,Z]] =0,\\
&\sum_{\text{cyclic}}[\nabla_\lambda, [\nabla_\mu, \nabla_\nu]] V^\rho =0 \qrq \nabla_{[\lambda} R^\rho{}_{|\sigma|\mu\nu]}=0,\\
&\sum_{\text{cyclic}} [\Delta_X, [\Delta_Y,\Delta_Z]] V^\mu =0 \qrq \sum_{\text{cyclic}}\big[ \Delta_X \mathcal{A}_{Y,Z} + \mathcal{A}_{X, [Y,Z]} \big] = 0,\\
&\sum_{\text{cyclic}}[\cl_{X},[\nabla_{\m},\nabla_{\n}]]V^\r=0 \qrq \cl_X R^\rho{}_{\sigma\mu\nu} = \nabla_\mu (\cl_X \Gamma^\rho_{\nu\sigma}) - (\m\leftrightarrow\n),\\
&\sum_{\text{cyclic}}[\Delta_{X},[\nabla_{\m},\nabla_{\n}]]V^\r=0 \qrq \Delta_X R^\rho{}_{\sigma\mu\nu} = \nabla_\mu (\Delta_X \Gamma^\rho_{\nu\sigma})+[\Delta_X,\p_\m]\Gamma^\rho_{\nu\sigma} - (\m\leftrightarrow\n),\\
&\sum_{\text{cyclic}}[\mathcal{L}_X, [\mathcal{L}_Y, \nabla_\m]] V^\r=0 \qrq \mathcal{L}_{[X,Y]} \Gamma^\rho_{\mu\nu} = [\mathcal{L}_X, \mathcal{L}_Y] \Gamma^\rho_{\mu\nu},
\end{align}
\end{small}
\end{subequations}
while the other four Jacobi identities yield no additional identities. Here $[\Delta_X,\p_\m]\Gamma^\rho_{\nu\sigma} $ is just a formal expression, and we do not know how to expand it. In the following coordinate-free formulation \eqref{C.2e}, the meaning is precise.

Now we discuss the Jacobi identity in the coordinate-free way. To be more general, the remaining formulas in this subsection allow a general affine connection with torsion $T$, while the main text otherwise specializes to the Levi-Civita connection. Similar to \eq{Jacobi ids}, we obtain
\begin{subequations}
\begin{small}
\begin{align}
&\sum_{\text{cyclic}}[\mathcal{L}_X, [\mathcal{L}_Y, \cl_Z]] V =0 \qrq \sum_{\text{cyclic}}\big[R(X,Y)Z-(\nabla_X T)(Y,Z)-T(T(X,Y),Z) \big]=0,\label{1stBianchi}\\
&\sum_{\text{cyclic}}[\nabla_X, [\nabla_Y, \nabla_Z]] V =0  \qrq \sum_{\text{cyclic}}\bigl[(\nabla_XR)(Y,Z)+R(T(X,Y),Z)\bigr]=0,\\
&\sum_{\text{cyclic}} [\Delta_X, [\Delta_Y,\Delta_Z]] V =0 \qrq \sum_{\text{cyclic}}\big[ \Delta_X \mathcal{A}_{Y,Z} + \mathcal{A}_{X, [Y,Z]}\big] = 0,\\
&\sum_{\text{cyclic}}[\cl_{X},[\nabla_{Y},\nabla_{Z}]]V=0 \qrq (\mathcal L_XR)(Y,Z)=(\d_\nabla P_X)(Y,Z),\\
&\sum_{\text{cyclic}}[\Delta_{X},[\nabla_Y,\nabla_Z]]V=0 \qrq (\widehat\Delta_X R)(Y,Z) = (\d_\nabla Q_X)(Y,Z),\label{C.2e}\\
&\sum_{\text{cyclic}}[\mathcal{L}_X, [\mathcal{L}_Y, \nabla_Z]] V=0 \qrq P_{[X,Y]} = \mathcal{L}_XP_Y- \mathcal{L}_YP_X,
\end{align}
\end{small}
\end{subequations}
where the covariant exterior differential is defined as
\begin{align}
(\d_\nabla P_X)(Y,Z)=(\nabla_YP_X)(Z)
-(\nabla_ZP_X)(Y)
+P_X(T(Y,Z)).
\end{align}
The first Bianchi identity in \eq{1stBianchi} comes from substituting the definition of the torsion
\begin{align}
T(X,Y)=\nabla_XY-\nabla_YX-[X,Y]
\end{align}
into the Jacobi identity of the Lie bracket
\begin{align}
\sum_{\text{cyclic}}[X,[Y,Z]]=0.
\end{align}

\bibliographystyle{JHEP}
\bibliography{biblio}
\end{document}